\documentclass[a4paper,10pt]{revtex4}
\usepackage{graphicx} 
\usepackage{amsmath} 
\usepackage{amssymb} 
\usepackage{bm} 
\usepackage{dcolumn}
\usepackage{color}
\usepackage{mathrsfs}
\usepackage{amsfonts}
\usepackage{varioref}
\usepackage{mathrsfs}
\usepackage{graphicx}
\usepackage{latexsym}
\usepackage{amsmath}
\usepackage{amssymb}
\usepackage{textcomp}
\usepackage{amsbsy}
\usepackage{graphics}
\usepackage{epstopdf}
\usepackage{color}
\usepackage[caption=false]{subfig}

\RequirePackage[colorlinks,citecolor=blue,urlcolor=magenta,linkcolor=blue]{hyperref}
\input epsf

\allowdisplaybreaks[4]
\begin{document}

\tolerance=5000

\title{From inflation to reheating and their dynamical stability analysis in Gauss-Bonnet gravity}

\author{Sergei~D.~Odintsov$^{1,2,3}$\,\thanks{odintsov@ieec.uab.es},
Tanmoy~Paul$^{4}$\,\thanks{pul.tnmy9@gmail.com}} \affiliation{
$^{1)}$ ICREA, Passeig Luis Companys, 23, 08010 Barcelona, Spain\\
$^{2)}$ Institute of Space Sciences (ICE, CSIC) C. Can Magrans s/n, 08193 Barcelona, Spain\\
$^{3)}$ Labaratory for Theoretical Cosmology, International Centre of Gravity and Cosmos,
Tomsk State University of Control Systems and Radioelectronics (TUSUR), 634050 Tomsk, Russia\\
$^{4)}$ National Institute of Technology Jamshedpur, Department of Physics, Jamshedpur - 831 014, India.}


\tolerance=5000

\begin{abstract}
We investigate the inflation and reheating phenomenology in scalar-Einstein-Gauss-Bonnet theory of gravity where a scalar field non-minimally couples with the Gauss-Bonnet (GB) curvature term. Regarding the inflationary phenomenology, we find -- (1) the inflation starts with a quasi de-Sitter phase and has an exit at a finite e-fold, which is indeed consistent with the resolution of horizon and flatness problems, (2) the scalar and tensor perturbations prove to be ghost free and do not suffer from gradient instability, (3) the curvature perturbation amplitude as well as its tilt and the tensor-to-scalar ratio turn out to be simultaneously compatible with the recent Planck data for suitable values of the parameters which also results to the inflationary energy scale $\sim 10^{12}\mathrm{GeV}$. After the inflation ends, the scalar field starts to decay to radiation with a constant decay width. For our considered scalar potential and the GB coupling function, the model results to an analytic power law solution of the Hubble parameter and a logarithmic solution of the scalar field during the reheating era, where the exponent of the Hubble parameter determines the effective EoS parameter ($w_\mathrm{eff}$) during the same. The stability of such reheating dynamics is examined by dynamical analysis which ensures that $w_\mathrm{eff}$ can go beyond unity and reach up-to the maximum value of $\mathrm{max}(w_\mathrm{eff}) = 1.56$. The scenario with $w_\mathrm{eff} > 1$ proves to be purely due to the presence of the GB coupling function, which in turn may have important consequences on enhancing the primordial gravitational waves' amplitude observed today. The inflationary e-fold number ($N_\mathrm{f}$) gets further constrained by the input of the reheating stage, and we critically examine the constraint of $N_\mathrm{f}$ coming from both the inflation and reheating phenomenology. We finally construct the complete forms of scalar potential ($V(\phi)$) and the GB coupling function ($\xi(\phi)$) that smoothly transits from inflation to reheating, and numerically solve the Hubble parameter and the scalar field for such complete forms of $V(\phi)$ and $\xi(\phi)$. Such numerical solutions match with analytic ones in the respective regime. After the reheating ends, the GB term gets decouple from the theory and the radiation energy density dominates the energy budget of the universe, which in turn produces the standard radiation era of the universe.
\end{abstract}

\maketitle

\section{Introduction}\label{SecI}
Today's cosmology is mostly driven by data which opens up the window to formulate a consistent theory of the universe that is on equal footing with the observations. In this light, it is intriguing that the current observations and experiments delimits the evolution of the universe in terms of the well established laws of physics, e.g. general relativity or some modified theories of gravity, the Standard Model of particle physic etc. However, direct experimental results probing physics above the TeV scale ceases to exist which also turns out to be a major impediment towards an unambiguous understanding of the physics of the very early universe. This gives rise to various early universe scenarios, namely the inflationary scenario \cite{guth,Linde:2005ht,Langlois:2004de,Riotto:2002yw,barrow1,barrow2,Baumann:2009ds}, the bouncing scenario \cite{Brandenberger:2012zb,Brandenberger:2016vhg,Battefeld:2014uga,Elizalde:2020zcb}, the emergent universe scenario \cite{Ellis:2003qz,Paul:2020bje,Li:2019laq} etc, all of which prove to be consistent with the Planck observations \cite{Akrami:2018odb}.

In the present work, we will take the route of inflation to describe the early phase of the universe, which elegantly resolves the horizon and flatness problems. The main essence of inflation is that the universe is dominated by constant energy density of a scalar field under consideration (the scalar field either minimally or non-minimally couples with the gravity sector), which in turn results to a quasi de-Sitter expansion of the universe. In this picture, the quantum fluctuations of the scalar field generates the primordial cosmological perturbation that are well compatible with the observational data \cite{Baumann:2009ds,Guth:1982ec,Abbott:1984fp}. Beside the curvature perturbation, the inflationary scenario generates the tensor perturbation as well, resulting to primordial gravitational waves which are of considerable interest now-a-days \cite{Baumann:2009ds,Boyle:2005se,Krauss:2010ty}. Being the spacetime curvature is large during inflation, it is legitimate to consider higher curvature corrections (over the usual Einstein-Hilbert term) in the gravitational action, generally known as higher curvature gravity theory. Such higher curvature corrections may originate from the diffeomorphism property of the action or from the low energy effective action of String theory. In fact, the first model of F(R) which remains viable up to date is the Starobinsky model \cite{Starobinsky:1982ee}, and ever since many models have been developed in various forms of modified gravity (see \cite{Nojiri:2017ncd,Nojiri:2010wj,Nojiri:2006ri,Capozziello:2011et,Capozziello:2010zz,delaCruzDombriz:2012xy,Olmo:2011uz} for reviews in modified gravity cosmology). In general, inclusion of higher curvature terms may lead to higher derivatives (higher than two) in the gravitational equations that hints to the Ostragradsky instability. However the scalar-Einsten-Gauss-Bonnet gravity theory is free from such instability due to the special appearance of Reimann tensor, Ricci tensor and Ricci scalar respectively. This is unlike to F(R) theory where, in general, the Ostragradsky instability appears. Actually a F(R) theory can be equivalently mapped to a minimally coupled scalar-tensor theory where the instability of the original Jordan frame is reflected through the kinetic term of the scalar field.

After the inflation ends, the universe passes through a non-trivial stage, namely the reheating stage when the inflaton decays to normal particles with a certain decay rate \cite{Martin:2010kz,Dai:2014jja,Cook:2015vqa,Chung:1998rq,Haque:2020zco,Saha:2020bis,Ellis:2021kad,Cai:2015soa,DeHaro:2017abf,Braden:2010wd,Nakayama:2008wy}. Reheating is one of the most important phases of the early universe. It essentially links our standard thermal universe with the pre-thermal state such as the inflationary universe through a complex non-linear process. Over the years, major cosmological observations have given us ample data to understand the theoretical as well as observational aspects of both the thermal and the very early non-thermal inflationary universe cosmology to an unprecedented level. However, our understanding of the intermediate reheating phase is still at a preliminary stage both in terms of both theory and observations. It is generically described by the coherently oscillating inflaton, and its non-linear decay into radiation. In the Boltzmann description, the reheating phase is parametrized by the reheating temperature ($T_\mathrm{re}$) and an effective equation of state (EoS) parameter ($w_\mathrm{eff}$). Until now, both these parameters remain relatively unconstrained, apart from the understanding that the reheating temperature is bounded by the BBN temperature, i.e $T_\mathrm{re} > T_\mathrm{BBN} \sim 10^{-2}\mathrm{GeV}$. As far as the EoS parameter is concerned, it is generally assumed that $w_\mathrm{eff}$ is time dependent during the decay of the inflaton to the radiation field. The evolution of $w_\mathrm{eff}$ must reach the value $\frac{1}{3}$ at the end of reheating in order to smoothly connect the reheating phase to the radiation dominated epoch. However, the initial value of $w_\mathrm{eff}$ (i.e at the start of reheating) is badly constrained and hence the dynamics of $w_\mathrm{eff}$ remains poorly understood. Here it deserves mentioning that some attempts have been made in order to indirectly probe the reheating EoS parameter particularly from the perspective of primordial gravitational waves or from inflationary magnetogenesis (see \cite{Kobayashi:2019uqs,Bamba:2020qdj,Haque:2021dha,Bamba:2021wyx}).

Motivated by the above arguments regarding the reheating stage, in the present work, we will investigate the inflation as well as the reheating phenomenology and try to constrain the $w_\mathrm{eff}$ in the context of ghost free scalar-Einstein-Gauss-Bonnet theory of gravity. The Gauss-Bonnet theory is well motivated due to its rich cosmological implications during early universe \cite{Nojiri:2005vv,Li:2007jm,Carter:2005fu,Nojiri:2019dwl,Odintsov:2020sqy,Cognola:2006eg,Oikonomou:2021kql,Odintsov:2018zhw,Chakraborty:2018scm,Kanti:2015pda,Pozdeeva:2021iwc,Vernov:2021hxo,Granda:2019jqy,Nojiri:2022xdo,Odintsov:2022unp}. Furthermore the holographic correspondence of such theory has been established in \cite{Nojiri:2020wmh}. In Gauss-Bonnet gravity theory, a scalar field non-minimally couples with the Gauss-Bonnet (GB) term, which makes the GB contribution non-trivial even in the four dimensional spacetime. The presence of the GB term affects the reheating era by two ways: (a) by implicit way through the dynamics of the scalar field, and (b) by explicit way through the derivatives of the GB coupling function. The scalar potential and the GB coupling function are considered to be of exponential types which allow analytic solutions of the field variables (i.e the Hubble parameter and the scalar field) during the inflation as well as during the reheating era. The stability of such dynamical evolutions are examined by dynamical analysis, which further puts a constraint on the reheating EoS parameter in the context of GB theory -- this may have important consequences on primordial gravitational waves (PGWs) as the effective EoS parameter during the reheating era has significant effects on enhancing the PGWs' spectrum. Thus in the current work, beside the inflationary and reheating phenomenology, the stability of reheating dynamics is also taken care of. Here we would like to mention that the reheating phenomenology in GB gravity theory has been performed earlier in a quite different context in \cite{Bhattacharjee:2016ohe,vandeBruck:2016xvt,Venikoudis:2022gfg} where the explicit effects of the GB term (via the derivatives of the GB coupling function) during the reheating stage has not been taken into account and also the stability of the reheating dynamics has not been examined. On contrary, in our present analysis, we will consider both the implicit and explicit effects of the GB term during the reheating stage, and examine the stability of the reheating dynamics by dynamical analysis -- these make our present work essentially different from earlier ones.

The paper is organized as follows: after a brief description of the considered scalar-Einstein-GB model in Sec.[\ref{sec-model}], we will describe the inflation and reheating phenomenology (along with the stability of the reheating dynamics) in Sec.[\ref{sec-inf}] and Sec.[\ref{sec-reh}] respectively. Finally we will establish the complete forms of scalar potential and GB coupling function that smoothly transits from inflation to reheating stage in Sec.[\ref{sec-complete}]. The paper ends with conclusions.

\section{The model}\label{sec-model}
We consider the scalar-Einstein-Gauss-Bonnet gravity theory with the following action:
\begin{eqnarray}
 S = \int d^4x \sqrt{-g}\left[\frac{R}{2\kappa^2} - \frac{1}{2}g^{\mu\nu}\partial_{\mu}\phi\partial_{\nu}\phi - V(\phi) + \frac{1}{2}\xi(\phi)\mathcal{G}\right]~~,
 \label{action-model}
\end{eqnarray}
where $R$ is the Ricci scalar formed by the metric $g_{\mu\nu}$, 
$\kappa = \sqrt{8\pi G}$ (with $G$ being the Newton's gravitational constant), $\phi$ is the scalar field under consideration embedded within the potential $V(\phi)$. Furthermore $\mathcal{G} = R^2 - 4R_{\mu\nu}R^{\mu\nu} + R_{\mu\nu\alpha\beta}R^{\mu\nu\alpha\beta}$ symbolizes the
Gauss-Bonnet (GB) scalar and $\xi(\phi)$ is the corresponding coupling between the GB and the scalar field. The gravitational and the scalar
field equations are obtained by varying the action with respect to $g_{\mu\nu}$ and $\phi$, and are given by,
\begin{eqnarray}
 \frac{1}{\kappa^2}&\big[&-R^{\mu\nu} + \frac{1}{2}Rg^{\mu\nu}\big] + \frac{1}{2}\partial^{\mu}\phi\partial^{\nu}\phi - \frac{1}{4}g^{\mu\nu}\partial^{\rho}\phi\partial_{\rho}\phi + \frac{1}{2}g^{\mu\nu}\left[V(\phi) - \xi(\phi)\mathcal{G}\right]\nonumber\\
 &+&2\xi(\phi)\left[RR^{\mu\nu} + 2R^{\mu}_{\rho}R^{\nu\rho} + R^{\mu\rho\sigma\tau}R^{\nu}_{\rho\sigma\tau} - 2R^{\mu\rho\nu\sigma}R_{\rho\sigma}\right] - 2R~\nabla^{\mu}\nabla^{\nu}\xi - 4\nabla^2\xi\left[R^{\mu\nu} - \frac{1}{2}Rg^{\mu\nu}\right]\nonumber\\
 &+&4R^{\nu\rho}~\nabla_{\rho}\nabla^{\mu}\xi + 4R^{\mu\rho}~\nabla_{\rho}\nabla^{\nu}\xi - 4\nabla^{\rho}\nabla^{\sigma}\xi\left[g^{\mu\nu}R^{\rho\sigma} + R^{\mu\rho\nu\sigma}\right] = 0
 \label{grav-eom}
\end{eqnarray}
and 
\begin{eqnarray}
 g^{\mu\nu}~\nabla_{\mu}\nabla_{\nu}\phi - \frac{\partial V}{\partial\phi} + \mathcal{G}\frac{\partial\xi}{\partial\phi} = 0
 \label{scalar-eom}
\end{eqnarray}
respectively. It may be noted that the field equations do not have higher than second order derivative of the metric. This is one of the advantages of the
Gauss-Bonnet gravity theory, comparing with the F(R) gravity where the gravitational equations generally contain more than second order derivative 
of the metric that eventually lead to some unstable situation. 

The spatially flat FLRW metric ansatz fits our purpose in the present context, i.e
\begin{eqnarray}
 ds^2 = -dt^2 + a^2(t)\delta_{ij}dx^{i}dx^{j}~~,
 \label{metric ansatz}
\end{eqnarray}
with $t$ designates the cosmic time, $a(t)$ being the scale factor of the universe and $\frac{\dot{a}}{a} = H$ will represent the Hubble parameter 
(an overdot shows the derivative $\frac{d}{dt}$). 
Due to this homogeneous and isotropic metric ansatz, the 
Ricci scalar and the GB scalar are given by: $R = 6\left(2H^2 + \dot{H}\right)$ and $\mathcal{G} = 24H^2\left(H^2 + \dot{H}\right)$ respectively. As a 
result along with the consideration that the scalar field is homogeneous in space, 
the temporal and the spatial component of Eq.(\ref{grav-eom}) take the following forms,
\begin{eqnarray}
 3H^2&=&\kappa^2\left[\frac{1}{2}\dot{\phi}^2 + V(\phi) - 12\dot{\xi}H^3\right]~~,\label{FRW-1}\\
 -2\dot{H}&=&\kappa^2\left[\dot{\phi}^2 + 4\ddot{\xi}H^2 + 4\dot{\xi}H\left(2\dot{H} - H^2\right)\right]~~,\label{FRW-2}
\end{eqnarray}
and moreover, the scalar field equation turns out to be,
\begin{eqnarray}
 \ddot{\phi} + 3H\dot{\phi} + V'(\phi) - 12\xi'(\phi)\left(H^4 + H^2\dot{H}\right) = 0~~.
 \label{scalar eom}
\end{eqnarray}
However the above three equations are not independent, one of them can be obtained from the other two. The important point to be mentioned is that 
the possible effects of GB term is reflected through derivative of $\xi(\phi)$ (with respect to either cosmic time or scalar field) in the field equations, 
which indicates that a constant coupling makes the GB contributions trivial. As mentioned earlier, we are mainly interested in describing the reheating 
dynamics and connect it to the inflationary era in the present context of GB theory. For this purpose, the scalar potential and the GB coupling function 
are considered to have the following forms:
\begin{eqnarray}
 V_\mathrm{I}(\phi) = V_0\mathrm{e}^{-\lambda\kappa\phi}~~~~~\mathrm{and}~~~~~\xi_\mathrm{I}(\phi) = \xi_0\mathrm{e}^{-\eta\kappa\phi}~,\label{inf-pot}
\end{eqnarray}
during inflation, and
\begin{eqnarray}
 V_\mathrm{R}(\phi) = V_1\mathrm{e}^{2\left(\phi-\phi_\mathrm{s}\right)/\phi_\mathrm{0}}
 ~~~~\mathrm{and}~~~~\xi_\mathrm{R}(\phi) = \xi_1\mathrm{e}^{-2\left(\phi - \phi_\mathrm{s}\right)/\phi_\mathrm{0}}~,\label{reh-pot}
\end{eqnarray}
during reheating era. Here both $V_0$ and $V_1$ have mass dimension [+4], while $\xi_0$ and $\xi_1$ are dimensionless parameters.
The respective forms of $V(\phi)$ and $\xi(\phi)$ are joined smoothly at the junction between inflation-to-reheating for suitable values of parameters, 
and their full forms that describe from the inflation to the reheating stage of the universe will be demonstrated in Sec.~[\ref{sec-complete}]. However before moving
to such an unified scenario, we first individually address the inflation and the reheating era, and investigate their stability condition in the 
context of GB gravity. The inflationary $V_\mathrm{I}(\phi)$ and $\xi_\mathrm{I}(\phi)$ in Eq.~(\ref{inf-pot}) are well motivated as they are consistent 
with the slow roll conditions, as we will see, and thus may produce a viable inflationary scenario. On other side, the reheating $V_\mathrm{R}(\phi)$ and 
$\xi_\mathrm{R}(\phi)$ in Eq.~(\ref{reh-pot}) are motivated as they lead to an analytic form of $H \sim t^{-1}$ which are suitable for the universe's 
evolution after inflation. Moreover due to such forms of $V_\mathrm{R}(\phi)$ and $\xi_\mathrm{R}(\phi)$, the reheating dynamics proves to be a 
stable attractor in the present context, as we will show in the subsequent sections.

After the end of reheating, the scalar field energy density transforms to relativistic particles which initiates the radiation dominated era with a suitable
reheating temperature (should be less than $\sim 10^{-2}\mathrm{GeV}$ or $\sim 10\mathrm{MeV}$ from the BBN constraint). As a result, 
the higher curvature GB term becomes a surface term in the four dimensional gravitational action after the reheating phase, by considering that the radiation energy does not exhibit any non-minimal coupling with the GB term, and thus the action may be written as,
\begin{eqnarray}
 S = \int d^4x \sqrt{-g}\left[\frac{R}{2\kappa^2} + \mathcal{L}_\mathrm{rad}\right]~~,\label{1}
\end{eqnarray}
where $\mathcal{L}_\mathrm{rad}$ represents the Lagrangian for the radiation content. This in turn 
results to the Standard Big-Bang Cosmology (SBBC) after the 
reheating era in the current context of scalar-Einstein-Gauss-Bonnet theory. Here it deserves mentioning that the speed of gravitational waves ($c_\mathrm{T}$) corresponds to the above action is unity. Thus after the reheating phase, the scalar-Einstein-GB gravity reduces to the standard Einstein gravity with a perfect fluid leading to $c_\mathrm{T} = 1$. This is indeed consistent with GW170817 event, according to which, the gravitational waves propagate with the speed of light (which is unity in natural unit).

\section{Inflationary era}\label{sec-inf}
\begin{figure}[!h]
\begin{center}
\centering
\includegraphics[width=3.0in,height=2.0in]{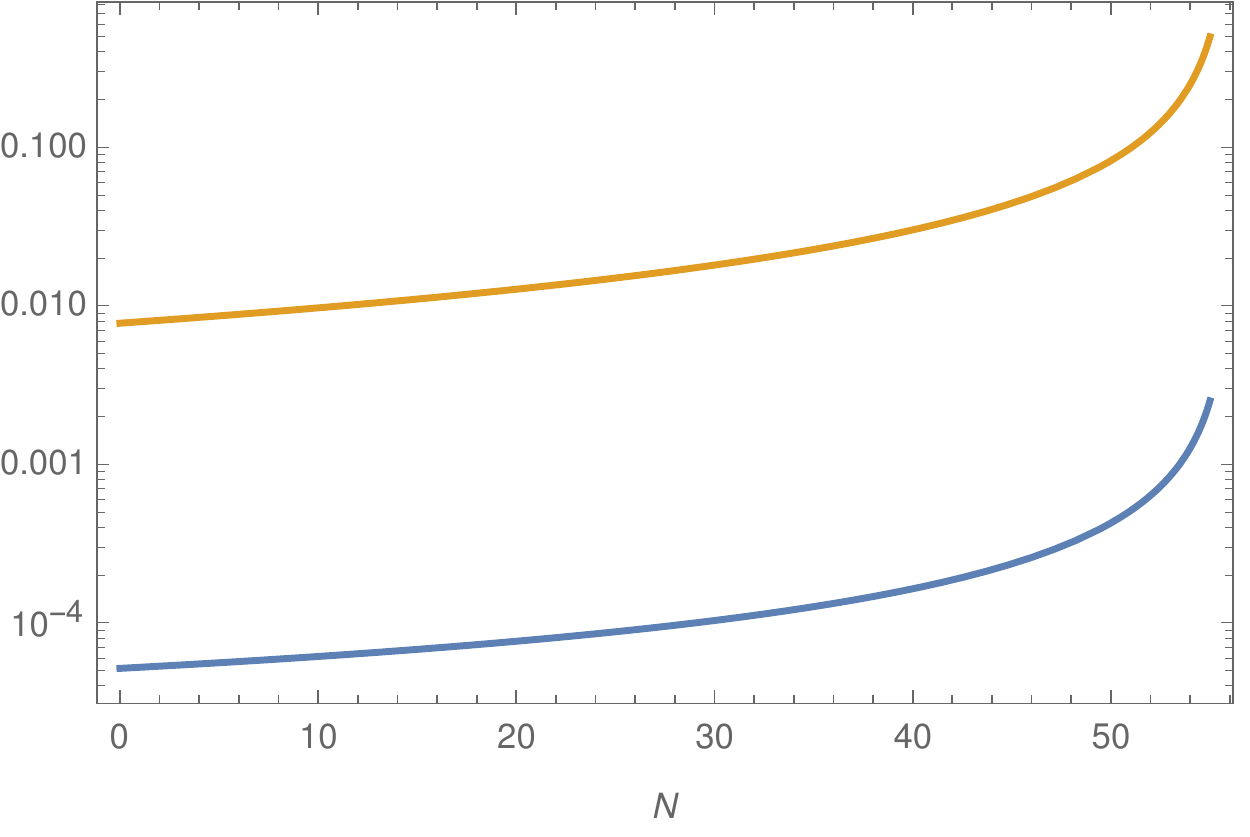}
\includegraphics[width=3.0in,height=2.0in]{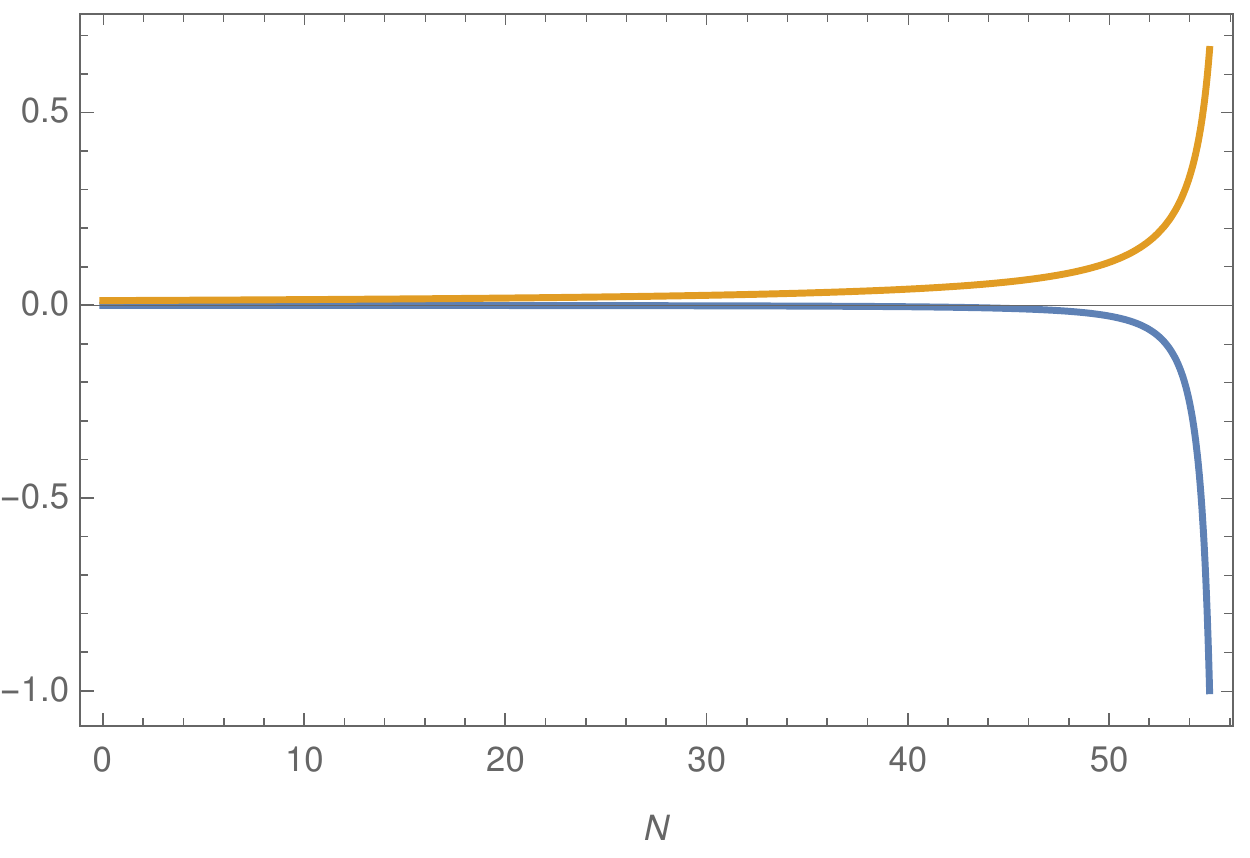}
\caption{{\color{blue}Left Plot}: $\epsilon_0$ vs. $N$ (blue curve) and $\epsilon_1$ vs. $N$ (yellow curve); {\color{blue}Right Plot}: $\delta_0$ vs. $N$ (blue curve) and $\delta_1$ vs. $N$ (yellow curve). In both the plots, we take $N_\mathrm{f} = 55$, and the parameter values are considered as: $\lambda = -0.005$, $\eta = 1$, $\kappa^4V_0 = 10^{-12}$, $\kappa^4V_0\xi_0 = 1$.}
\label{plot-SR}
\end{center}
\end{figure}

In the inflationary era, the universe expands through a quasi de-Sitter stage which is best described by slowness of the following parameters 
(known as slow-roll parameters) in scalar-Einstein-GB theory \cite{Guo:2010jr},
\begin{eqnarray}
 \epsilon_0&=&-\dot{H}/H^2~~~~~~~~~~,~~~~~~~~~~~\epsilon_1 = \frac{\dot{\epsilon}_0}{H\epsilon_0}~~,\nonumber\\
 \delta_0&=&-4\dot{\xi}H~~~~~~~~~~~~,~~~~~~~~~~~~\delta_1 = \frac{\dot{\delta}_0}{H\delta_0}~.
 \label{Sr quantities}
\end{eqnarray}
The slow-roll parameters are written in the unit of $\kappa = 1$; in fact, all the subsequent calculations are based in this unit. The first two slow-roll parameters, i.e $\epsilon_0$ and $\epsilon_1$, are usual in the minimally coupled scalar-tensor theory, while $\delta_i$ are
defined due to the existence of the non-minimal GB coupling. 
The conditions $\epsilon_i \ll 1$ and $\delta_i \ll 1$ ensure a negligible exchange of energy between the scalar field and the gravity sector, which in turn
leads to a quasi de-Sitter evolution of the universe. With these slow roll conditions, the field equations turn out to be,  
\begin{eqnarray}
 0&=&-3H^2 + V_\mathrm{I}(\phi)~~,\nonumber\\
 0&=&2\dot{H} + \left[\dot{\phi}^2 - 4\dot{\xi}_\mathrm{I}H^3\right]~~,\nonumber\\
 0&=&3H\dot{\phi} + V_\mathrm{I}'(\phi) - 12\xi_\mathrm{I}'(\phi)H^4~~.
 \label{SR eom}
\end{eqnarray}
The above slow roll equations immediately lead to the slow roll quantities $\epsilon_0$ and $\delta_0$, in terms of $V_\mathrm{I}(\phi)$ and $\xi_\mathrm{I}(\phi)$, as follows:
\begin{eqnarray}
 \epsilon_0&=&\frac{1}{2}\left\{\left(\frac{V_\mathrm{I}'}{V_\mathrm{I}}\right)^2 - \frac{4}{3}\xi_\mathrm{I}'(\phi)V_\mathrm{I}'(\phi)\right\}~~,\nonumber\\
 \delta_0&=&\frac{4}{3}\left\{\xi_\mathrm{I}'(\phi)V_\mathrm{I}'(\phi) - \frac{4}{3}V_\mathrm{I}^2\left(\xi_\mathrm{I}'(\phi)\right)^2\right\}~~,
 \label{SR intermediate}
\end{eqnarray}
respectively, where the overprime with the argument $\phi$ symbolizes $\frac{d}{d\phi}$. By using the explicit forms of $V_\mathrm{I}(\phi)$ and $\xi_\mathrm{I}(\phi)$ from Eq.~(\ref{inf-pot}) along with a little bit
of simplification, we get the final expressions of $\epsilon_0$ and $\delta_0$ as,
\begin{eqnarray}
 \epsilon_0&=&\frac{\lambda}{6}\left\{3\lambda - 4V_0\xi_0\eta\mathrm{e}^{-\left(\lambda + \eta\right)\phi}\right\}~~,\nonumber\\
 \delta_0&=&\frac{4}{9}V_0\xi_0\eta\left\{3\lambda\mathrm{e}^{\left(\lambda + \eta\right)\phi} - 4V_0\xi_0\eta\right\}
 \mathrm{e}^{-2\left(\lambda + \eta\right)\phi}~.\label{SR-1}
\end{eqnarray}
Consequently one may determine,
\begin{eqnarray}
 \epsilon_1&=&\frac{4}{3}V_0\xi_0\eta\left(\lambda + \eta\right)\mathrm{e}^{-\left(\lambda + \eta\right)\phi}~~,\nonumber\\
 \delta_1&=&-\frac{1}{3}\left(\lambda + \eta\right)\left\{3\lambda - 8V_0\xi_0\eta\mathrm{e}^{-\left(\lambda + \eta\right)\phi}\right\}~~.
 \label{SR-2}
\end{eqnarray}
From Eq.~(\ref{SR-1}) and Eq.~(\ref{SR-2}), it is necessary to ensure that $\epsilon_i \ll 1$ and $\delta_i \ll 1$ during the inflationary era. For this purpose, we need
the viable constraints on $\lambda$ and $\eta$ based on the consistency of the theoretical expectations of the model with the recent Planck 2018 observation \cite{Akrami:2018odb}. In the present context, we consider the end of inflation to be the instant of $\delta_0 = -1$. On contrary, if we take the condition $\epsilon_0 = 1$ to be the end of inflation, then the slow roll
parameters like $\epsilon_1$, $\left|\delta_0\right|$ and $\delta_1$ seem to be much larger than unity during the inflationary epoch. This breaks the slow roll condition. In order
to circumvent this problem, here we consider $\left|\delta_0\right| = 1$ to be the end of inflation, in which case, the other slow roll parameters
remain less than unity during the inflation and validates the slow roll conditions. In the context of scalar-Einstein-GB gravity, the condition for the end of inflation other than $\epsilon_0 = 1$ has been considered in literature, see \cite{Odintsov:2018zhw,Granda:2019jqy}. Due to Eq.~(\ref{SR-1}), the $\delta_0 = -1$ leads to the following algebraic equation for
$\phi_\mathrm{f}$ (the subscript 'f' denotes at the end of inflation):
\begin{eqnarray}
 \mathrm{e}^{2\left(\lambda + \eta\right)\phi_\mathrm{f}} + \frac{4}{3}V_0\xi_0\eta\lambda\mathrm{e}^{\left(\lambda + \eta\right)\phi_\mathrm{f}} 
 - \frac{16}{9}\left(V_0\xi_0\eta\right)^2 = 0\nonumber
\end{eqnarray}
which can be solved for $\phi_\mathrm{f}$ to get,
\begin{eqnarray}
 \phi_\mathrm{f} = \frac{1}{\left(\lambda + \eta\right)}\ln{\left[\frac{2}{3}V_0\xi_0\eta\left\{\sqrt{\left(\lambda^2 + 4\right)} - \lambda\right\}\right]}~~.
 \label{end phi}
 \end{eqnarray}
By using the above expression, we can determine the inflationary e-folding number ($N_\mathrm{f}$) that is defined as, 
 \begin{eqnarray}
  N_\mathrm{f} = \int_{\phi_\mathrm{i}}^{\phi_\mathrm{f}}\left(\frac{H}{\dot{\phi}}\right)d\phi = 
  \int_{\phi_\mathrm{i}}^{\phi_\mathrm{f}}\left(\frac{-V_\mathrm{I}(\phi)}{V_\mathrm{I}'(\phi) - \frac{4}{3}V_\mathrm{I}^2(\phi)\xi_\mathrm{I}'(\phi)}\right)d\phi~~,
  \label{e-fold-1}
 \end{eqnarray}
where $\phi_\mathrm{i}$ and $\phi_\mathrm{f}$ represent the scalar field at the beginning of inflation and at the end of inflation respectively. The 
beginning of inflation is considered to be the instance when the CMB scale mode ($\sim 0.05 \mathrm{Mpc}^{-1} \sim 10^{-40}\mathrm{GeV}$) crosses 
the horizon. Moreover the second equality 
in the above expression is obtained by using the slow roll field Eq.~(\ref{SR eom}). Due to $V_\mathrm{I}(\phi) = V_0\mathrm{e}^{-\lambda\phi}$ and 
$\xi_\mathrm{I}(\phi) = \xi_0\mathrm{e}^{-\eta\phi}$, the integration of Eq.(\ref{e-fold-1}) is performed to get the following result,
 \begin{eqnarray}
  N_\mathrm{f} = \frac{1}{\lambda\left(\lambda + \eta\right)}
  \ln{\left[\frac{3\lambda\mathrm{e}^{\left(\lambda + \eta\right)\phi_\mathrm{f}} - 4V_0\xi_0\eta}
  {3\lambda\mathrm{e}^{\left(\lambda + \eta\right)\phi_\mathrm{i}} - 4V_0\xi_0\eta}\right]}~~.
  \label{e-fold-2}
 \end{eqnarray}
 Eq.~(\ref{e-fold-2}) may be inverted to obtain the initial scalar field in terms of $N_\mathrm{f}$ as, 
 \begin{eqnarray}
  \phi_\mathrm{i} = \frac{1}{\left(\lambda + \eta\right)}
  \ln{\left[\frac{V_0\xi_0\eta}{3\lambda}\left\{2\mathrm{e}^{-\lambda\left(\lambda + \eta\right)N_\mathrm{f}}\left(\lambda\sqrt{\left(\lambda^2 + 4\right)} - \lambda^2 - 2\right) + 4\right\}\right]}~~,
  \label{start phi}
 \end{eqnarray}
 where we use the $\phi_\mathrm{f}$ from Eq.~(\ref{end phi}). Having set the stage, we now demonstrate the evolution of the slow roll parameters that have been determined 
 in Eq.~(\ref{SR-1}) and Eq.~(\ref{SR-2}), where $\phi$ is the scalar field at $N$ e-folds of inflation, i.e 
 $N = \int_{\phi_\mathrm{i}}^{\phi(N)}\left(\frac{H}{\dot{\phi}}\right)d\phi$ (clearly the inflation starts from $N = 0$). Consequently 
 Eq.~(\ref{start phi}) and Eq.~(\ref{end phi}) lead to the evolution of the scalar field during inflation, in particular $\phi = \phi(N)$, as,
 \begin{eqnarray}
  \phi(N) = \frac{1}{\left(\lambda + \eta\right)}
  \ln{\left[\frac{V_0\xi_0\eta}{3\lambda}\left\{2\mathrm{e}^{-\lambda\left(\lambda + \eta\right)\left(N_\mathrm{f} - N\right)}
  \left(\lambda\sqrt{\left(\lambda^2 + 4\right)} - \lambda^2 - 2\right) + 4\right\}\right]}~~.
  \label{e-fold-4}
 \end{eqnarray}
 Owing to the above solution of $\phi(N)$, Eq.~(\ref{SR eom}) immediately leads to the Hubble parameter as,
 \begin{eqnarray}
  H(N) = \left(\frac{V_0}{3}\right)^{\frac{1}{2}}
  \left[\frac{V_0\xi_0\eta}{3\lambda}\left\{2\mathrm{e}^{-\lambda\left(\lambda + \eta\right)\left(N_\mathrm{f} - N\right)}
  \left(\lambda\sqrt{\left(\lambda^2 + 4\right)} - \lambda^2 - 2\right) + 4\right\}\right]^{\frac{-\lambda}{2\left(\lambda + \eta\right)}}~~.
  \label{Hubble-evolution}
 \end{eqnarray}
Thus as a whole, Eq.~(\ref{e-fold-4}) and Eq.~(\ref{Hubble-evolution}) represent the evolution of the scalar field and of the Hubble parameter during inflation respectively. By using Eq.~(\ref{e-fold-4}), we give the plots of the slow roll parameters during the inflationary era from Eq.~(\ref{SR-1}) and Eq.~(\ref{SR-2}), see
 Fig.~[\ref{plot-SR}] where we take $N_\mathrm{f} = 55$ and the parameters are considered to be: $\lambda = -0.005$, $\eta = 1$, $\kappa^4V_0 = 10^{-12}$ and $\kappa^4V_0\xi_0 = 1$ which are well consistent with the Planck data as we will show in the following subsection (recall that $\kappa^{-1} = \frac{1}{\sqrt{8\pi G}} \approx 10^{19}\mathrm{GeV}$ with $G$ being the Newton's gravitational constant, and moreover $\xi_0$ is dimensionless). Fig.[\ref{plot-SR}] clearly demonstrates the following points: (1) the slow roll parameter $\delta_0$ is negative and acquires the value $= -1$ at $N = N_\mathrm{f}$, i.e at the end of inflation -- as expected. The negative values of $\delta_0$ is due to the fact that $\frac{d\phi}{dN} < 0$ during inflation. (2) $\epsilon_0$, $\epsilon_1$, $\left|\delta_0\right|$ and $\delta_1$ remain less than unity for the whole inflationary stage, which in turn validates the slow roll conditions. Here it deserves mentioning that the slow roll conditions in the present context get violated if one considers $\epsilon_0 = 1$ to be the end point of inflation, instead of $\delta_0 = -1$. This is the reason that $\delta_0 = -1$ becomes the suitable condition for the exit of inflation.

 
 
\subsection*{Cosmological phenomenology}

We begin by defining the functions $Q_\mathrm{i}$ (see \cite{Hwang:2005hb,Noh:2001ia} for details) as follows:
\begin{eqnarray}
 Q_\mathrm{a}&=&4\dot{\xi}H^2~~~~~~~~~~~,~~~~~~~~~~Q_\mathrm{b} = 8\dot{\xi}H~~~~~~~,~~~~~~~~~Q_\mathrm{c} = 0~~,\nonumber\\
 Q_\mathrm{d}&=&0~~~~~~~~~~~~~~~~,~~~~~~~~~~~Q_\mathrm{e} = 16\dot{\xi}\dot{H}~~~~~~~,~~~~~~~Q_\mathrm{f} = -8\left(\ddot{\xi} - \dot{\xi}H\right)~~,
 \label{Q-s}
\end{eqnarray}
where the overdot represents $\frac{d}{dt}$.\\
The scalar and tensor perturbation over the spatially flat FRW background metric (\ref{metric ansatz}) are defined as:
\begin{eqnarray}
 ds^2 = -\left(1 + 2\psi\right)dt^2 + a^2(t)\left(1 - 2\psi\right)\delta_{ij}dx^{i}dx^{j}~~,\label{sp}
\end{eqnarray}
and
\begin{eqnarray}
 ds^2 = -dt^2 + a^2(t)\left(\delta_{ij} + h_\mathrm{ij}\right)dx^{i}dx^{j}~~,\label{tp}
\end{eqnarray}
where $\psi(t,\vec{x})$ and $h_\mathrm{ij}(t,\vec{x})$ represent the scalar and tensor perturbation variable respectively. In the comoving gauge where the velocity perturbation is taken to be zero, the scalar perturbation $\psi$ in Eq.~(\ref{sp}) resembles to the curvature perturbation, i.e $\mathcal{R} = \psi$. Thus we can work with the variable $\psi(t,\vec{x})$. The Fourier modes of the curvature perturbation, at linear order in perturbation theory, satisfies the following Mukhanov-Sasaki equation:
\begin{eqnarray}
 v_\mathrm{s}''(k,\tau) + \left(c_\mathrm{s}^2k^2 - \frac{z_\mathrm{s}''}{z_\mathrm{s}}\right)v_\mathrm{s}(k,\tau) = 0~~,
 \label{cp-1}
\end{eqnarray}
where $v_\mathrm{s}(k,\tau) = z_\mathrm{s}\mathcal{R}$ is the Mukhanov-Sasaki variable for the curvature perturbation, $k$ is the momentum of the perturbation mode and $\tau$ is the conformal time defined by $d\tau = \frac{dt}{a(t)}$ (please note that throughout the paper, we use $\tau$ as the proper time while $\eta$ is a model parameter present in the GB coupling function, see Eq.(\ref{inf-pot})). Moreover $z_\mathrm{s}$ and $c_\mathrm{s}$ (the speed for the curvature perturbation) in the present context of scalar-Einstein-GB gravity theory are given by \cite{Hwang:2005hb,Noh:2001ia},
\begin{eqnarray}
 z_\mathrm{s}^2 = a^2\left[\frac{\dot{\phi}^2 + 3Q_\mathrm{a}^2/\left(2 + Q_\mathrm{b}\right)}{\left\{H + Q_\mathrm{a}/\left(2 + Q_\mathrm{b}\right)\right\}^2}\right]~~~~~~~~~~~,~~~~~~~~~~~
 c_\mathrm{s}^2 = 1 + \left[\frac{\left(\frac{Q_\mathrm{a}}{2+Q_\mathrm{b}}\right)Q_\mathrm{e} + \left(\frac{Q_\mathrm{a}}{2+Q_\mathrm{b}}\right)^2Q_\mathrm{f}}
 {\dot{\phi}^2 + 3Q_\mathrm{a}^2/\left(2 + Q_\mathrm{b}\right)}\right]~~.
 \label{cp-2}
\end{eqnarray}
Using the forms of $Q_\mathrm{i}$ from Eq.(\ref{Q-s}), the above expressions of $z_\mathrm{s}$ and $c_\mathrm{s}$ turn out to be,
\begin{eqnarray}
 z_\mathrm{s}^2 = a^2\left[\frac{\dot{\phi}^2 - 6\Delta\dot{\xi}H^3}{H^2\left(1 - \frac{\Delta}{2}\right)^2}\right]~~~~~~~~~~~~\mathrm{and}~~~~~~~~~~~
 c_\mathrm{s}^2 = 1 - \frac{8\Delta\dot{\xi}H\dot{H} + 2\Delta^2H^2\left(\ddot{\xi} - \dot{\xi}H\right)}{\dot{\phi}^2 - 6\Delta\dot{\xi}H^3}~~,
 \label{cp-3}
\end{eqnarray}
respectively, where $\Delta = \frac{\delta_0}{1-\delta_0}$ (recall that $\delta_0 = -4\dot{\xi}H$ is the slow roll parameter). The slow roll Eq.(\ref{SR eom}), gives $\dot{\phi}^2 = -2\dot{H} + 4\dot{\xi}H^3$, by using which in the above expression of $z_\mathrm{s}$, one gets \cite{Guo:2010jr}
\begin{eqnarray}
 z_\mathrm{s}^2 = a^2\bigg\{\left(2\epsilon_0 - \delta_0\right) + \mathrm{higher~order~in~slow~roll~parameters}\bigg\}~~,
 \label{cp-4}
\end{eqnarray}
where the last term contains more than linear order of slow roll parameters. In order to solve the Mukhanov-Sasaki Eq.(\ref{cp-1}), we calculate $\frac{z_\mathrm{s}''}{z_\mathrm{s}}$. In particular, we get:
\begin{eqnarray}
 \frac{z_\mathrm{s}''}{z_\mathrm{s}} = a^2H^2\left\{2 -\epsilon_0 + \frac{3\dot{F}}{2HF} - \left(\frac{\dot{F}}{2HF}\right)^2 + \frac{\ddot{F}}{2HF} + \mathrm{H.O.~in~slow~roll~parameters}\right\}~~.
 \label{cp-5}
\end{eqnarray}
Here $F = 2\epsilon_0 - \delta_0$, and thus
\begin{eqnarray}
 \frac{\dot{F}}{HF} = \frac{2\epsilon_0\epsilon_1 - \delta_0\delta_1}{2\epsilon_0 - \delta_0}~~.
 \label{cp-6}
\end{eqnarray}
Eq.(\ref{cp-6}) clearly indicates that $\dot{F}/(HF)$ is of the linear order in slow roll parameter, and hence $\left(\frac{\dot{F}}{HF}\right)^2$ or $\frac{\ddot{F}}{HF}$ (present in Eq.(\ref{cp-5})) contains higher order or higher derivatives of the slow roll parameters respectively. Thereby plugging the expression of $\dot{F}/(HF)$ from Eq.(\ref{cp-6}) to Eq.(\ref{cp-5}) yields the following form of $\frac{z_\mathrm{s}''}{z_\mathrm{s}}$ as:
\begin{eqnarray}
 \frac{z_\mathrm{s}''}{z_\mathrm{s}} = a^2H^2\left\{2 - \epsilon_0 + \frac{3}{2}\left(\frac{2\epsilon_0\epsilon_1 - \delta_0\delta_1}{2\epsilon_0 - \delta_0}\right) + \mathrm{H.O.~in~slow~roll~parameters}\right\} ~~.
 \label{cp-7}
\end{eqnarray}
Furthermore, owing to the slow roll inflationary expansion in the background spacetime, we may write $aH = -\frac{1}{\tau\left(1-\epsilon_0\right)}$. This immediately results to the final form of $\frac{z_\mathrm{s}''}{z_\mathrm{s}}$ as follows:
\begin{eqnarray}
 \frac{z_\mathrm{s}''}{z_\mathrm{s}} = \frac{1}{\tau^2}\left\{2 + 3\epsilon_0 + \frac{3}{2}\left(\frac{2\epsilon_0\epsilon_1 - \delta_0\delta_1}{2\epsilon_0 - \delta_0}\right) + \mathrm{H.O.~in~slow~roll~parameters}\right\}~~.
 \label{cp-8}
\end{eqnarray}
Coming to $c_\mathrm{s}^2$, the term $\ddot{\xi}$ (present in the expression of $c_\mathrm{s}^2$) is given by (in terms of the slow roll parameters):
\begin{eqnarray}
 \ddot{\xi} = -\frac{\delta_0}{4}\left(\epsilon_0 + \delta_1\right)~~,
 \label{cp-9}
\end{eqnarray}
where we use $\dot{\xi} = -\delta_0/(4H)$. Due to the above Eq.(\ref{cp-9}) along with the slow roll evolution of $\dot{\phi}^2$, the $c_\mathrm{s}^2$ from Eq.(\ref{cp-2}) takes the following form:
\begin{eqnarray}
 c_\mathrm{s}^2 = 1 - \Delta^2\left\{\frac{2\epsilon_0 + \frac{\delta_0}{2}\left(1 - 5\epsilon_0 - \delta_1\right)}{2\epsilon_0 - \delta_0 + \frac{3}{2}\Delta\delta_0}\right\}~~.
 \label{cp-10}
\end{eqnarray}
Here we may recall that $\Delta = \delta_0/(1 - \delta_0)$. Thus as a whole, Eq.(\ref{cp-8}) and Eq.(\ref{cp-10}) provide the basic ingredients to solve the curvature perturbation Mukhanov-Sasaki equation. Under the slow roll condition, the quantities $\tau^2\frac{z_\mathrm{s}''}{z_\mathrm{s}}$ and $c_\mathrm{s}^2$ may be approximated to be constant. Then the general solution of Eq.(\ref{cp-1}) can be written as,
\begin{eqnarray}
 v_\mathrm{s}(k,\tau) = \frac{\sqrt{\pi|\tau|}}{2}\mathrm{e}^{i\pi(1+2\nu_s)/4}\left\{c_1~H_\mathrm{\nu_s}^{(1)}(c_\mathrm{s}k|\tau|) + c_2~H_\mathrm{\nu_s}^{(2)}(c_\mathrm{s}k|\tau|)\right\}~~,
 \label{cp-11}
\end{eqnarray}
with $H_\mathrm{\nu_s}^{(1)}(z)$ and $H_\mathrm{\nu_s}^{(2)}(z)$ are the Hankel functions of order $\nu_s$ having first and second kind respectively, and
\begin{eqnarray}
 \nu_s^2 = \frac{1}{4} + \left\{2 + 3\epsilon_0 + \frac{3}{2}\left(\frac{2\epsilon_0\epsilon_1 - \delta_0\delta_1}{2\epsilon_0 - \delta_0}\right)\right\}~~,
 \label{cp-12}
\end{eqnarray}
(by neglecting the higher order contributions of slow roll parameters). Moreover $c_1$ and $c_2$ are integration constants which are taken as $c_1 = 1$ and $c_2 = 0$ respectively, in order to have the Bunch-Davies vacuum of $v_\mathrm{s}(k,\tau)$ at asymptotic past (in particular, at $c_\mathrm{s}k|\tau| \rightarrow \infty$). Owing to this, the curvature perturbation power spectrum at super-Hubble regime comes as \cite{Guo:2010jr},
\begin{eqnarray}
 \mathcal{P}_\mathrm{s}(k,\tau) = \left(\frac{H^2}{4\pi^2}\right)\left(\frac{1}{aH|\tau|}\right)^2\left(\frac{1}{c_\mathrm{s}^3\left(2\epsilon_0 - \delta_0\right)}\right)\frac{\Gamma^2(\nu_s)}{\Gamma^2(3/2)}\left(\frac{c_\mathrm{s}k|\tau|}{2}\right)^{3-2\nu_s}~~.
 \label{cp-13}
\end{eqnarray}

For the tensor perturbation, the Fourier tensor perturbation Mukhanov-Sasaki variable is defined by $v_\mathrm{T}(k,\tau) = z_\mathrm{T}(\tau)h(k,\tau)$ where $h(k,\tau)$ is the Fourier version of $h_{ij}$ and $z_\mathrm{T}$ is given by \cite{Hwang:2005hb,Noh:2001ia,Guo:2010jr},
\begin{eqnarray}
 z_\mathrm{T}^2 = a^2\left\{1 + 4\dot{\xi}H\right\}~~.
 \label{tp-1}
\end{eqnarray}
The variable $v_\mathrm{T}(k,\tau)$ evolves through the following equation:
\begin{eqnarray}
 v_\mathrm{T}''(k,\tau) + \left(c_\mathrm{T}^2k^2 - \frac{z_\mathrm{T}''}{z_\mathrm{T}}\right)v_\mathrm{T}(k,\tau) = 0~~,
 \label{tp-2}
\end{eqnarray}
with $z_\mathrm{T}(\tau)$ is shown above and (see \cite{Hwang:2005hb,Noh:2001ia,Guo:2010jr})
\begin{eqnarray}
 c_\mathrm{T}^2 = 1 + \frac{4\left(\ddot{\xi} - \dot{\xi}H\right)}{1 + 4\dot{\xi}H}~~.
 \label{tp-3}
\end{eqnarray}
Here it may be mentioned that Eq.(\ref{tp-3}) indicates that the speed of the gravitational waves in the present context of scalar-Einstein-GB gravity theory is not unity, and the deviation of $c_\mathrm{T}^2$ from unity is controlled by the derivatives of the GB coupling function. However as demonstrated at the end of Sec.[\ref{sec-model}] that after the reheating phase, the scalar-Einstein-GB gravity of the current context reduces to the standard Einstein gravity with a perfect fluid (actually the scalar field energy density transforms to radiation energy during the reheating phase), in which case, the speed of the gravitational waves is unity. This is indeed consistent with GW170817 event, according to which, the gravitational waves' speed at $present~epoch$ of the universe is equal to the speed of light (which is unity in natural unit). Eq.(\ref{tp-1}) helps to calculate $\frac{z_\mathrm{T}''}{z_\mathrm{T}}$ (one of the main ingredients in the tensor Mukhanov-Sasaki equation), and it is given by,
\begin{eqnarray}
 \frac{z_\mathrm{T}''}{z_\mathrm{T}} = a^2H^2\left\{2 - \epsilon_0 + \frac{3\dot{Q}}{2HQ} - \left(\frac{\dot{Q}}{2HQ}\right)^2 + \frac{\ddot{Q}}{2HQ}\right\}~~.
 \label{tp-4}
\end{eqnarray}
Here $Q = 1-\delta_0$, and thus $\frac{\dot{Q}}{HQ} = -\delta_0\delta_1/(1-\delta_0)$ which is of second order in slow roll parameter. As a result, Eq.(\ref{tp-4}) can be expressed as,
\begin{eqnarray}
 \frac{z_\mathrm{T}''}{z_\mathrm{T}} = a^2H^2\bigg\{(2 - \epsilon_0) + \mathrm{H.O.~in~slow~roll~parameters}\bigg\}~~,
 \label{tp-5}
\end{eqnarray}
which, due to $aH = -\frac{1}{\tau\left(1-\epsilon_0\right)}$ during the slow roll inflationary evolution, is equivalently written as,
\begin{eqnarray}
 \frac{z_\mathrm{T}''}{z_\mathrm{T}} = \frac{1}{\tau^2}\bigg\{(2 + 3\epsilon_0) + \mathrm{H.O.~in~slow~roll~parameters}\bigg\}~~.
 \label{tp-6}
\end{eqnarray}
Coming to $c_\mathrm{T}^2$, plugging the expression of $\ddot{\xi}$ from Eq.(\ref{cp-9}) to Eq.(\ref{tp-3}) yields
\begin{eqnarray}
 c_\mathrm{T}^2 = (1 + \delta_0) + \mathrm{H.O.~in~slow~roll~parameters}~~.
 \label{tp-7}
\end{eqnarray}
Being $\tau^2\frac{z_\mathrm{T}''}{z_\mathrm{T}}$ and $c_\mathrm{T}^2$ depend on the slow roll quantities, we may consider $\frac{z_\mathrm{T}''}{z_\mathrm{T}} \propto \frac{1}{\tau^2}$ and $c_\mathrm{T}^2$ to be approximately constant during the inflation. As a result, the tensor Mukhanov-Sasaki Eq.(\ref{tp-2}) has the following solution:
\begin{eqnarray}
 v_\mathrm{T}(k,\tau) = \frac{\sqrt{\pi|\tau|}}{2}\mathrm{e}^{i\pi(1+2\nu_T)/4}\left\{H_\mathrm{\nu_T}^{(1)}(c_\mathrm{T}k|\tau|)\right\}~~,
 \label{tp-8}
\end{eqnarray}
with $\nu_T^2 = \frac{1}{4} + \left(2 + 3\epsilon_0\right)$ (by neglecting the higher order contributions of slow roll parameters). The above solution of $v_T(k,\tau)$ is indeed consistent with the Bunch-Davies vacuum at asymptotic past. Due to the above solution of $v_\mathrm{T}(k,\tau)$, the tensor perturbation power spectrum at super-Hubble regime turns out to be \cite{Guo:2010jr},
\begin{eqnarray}
 \mathcal{P}_\mathrm{T}(k,\tau) = \left(\frac{H^2}{4\pi^2}\right)\left(\frac{1}{aH|\tau|}\right)^2\left(\frac{8}{c_\mathrm{T}^3\left(1 - \delta_0\right)}\right)\frac{\Gamma^2(\nu_T)}{\Gamma^2(3/2)}\left(\frac{c_\mathrm{T}k|\tau|}{2}\right)^{3-2\nu_T}~~.
 \label{tp-9}
\end{eqnarray}

\begin{figure}[!h]
\begin{center}
\centering
\includegraphics[width=3.0in,height=2.0in]{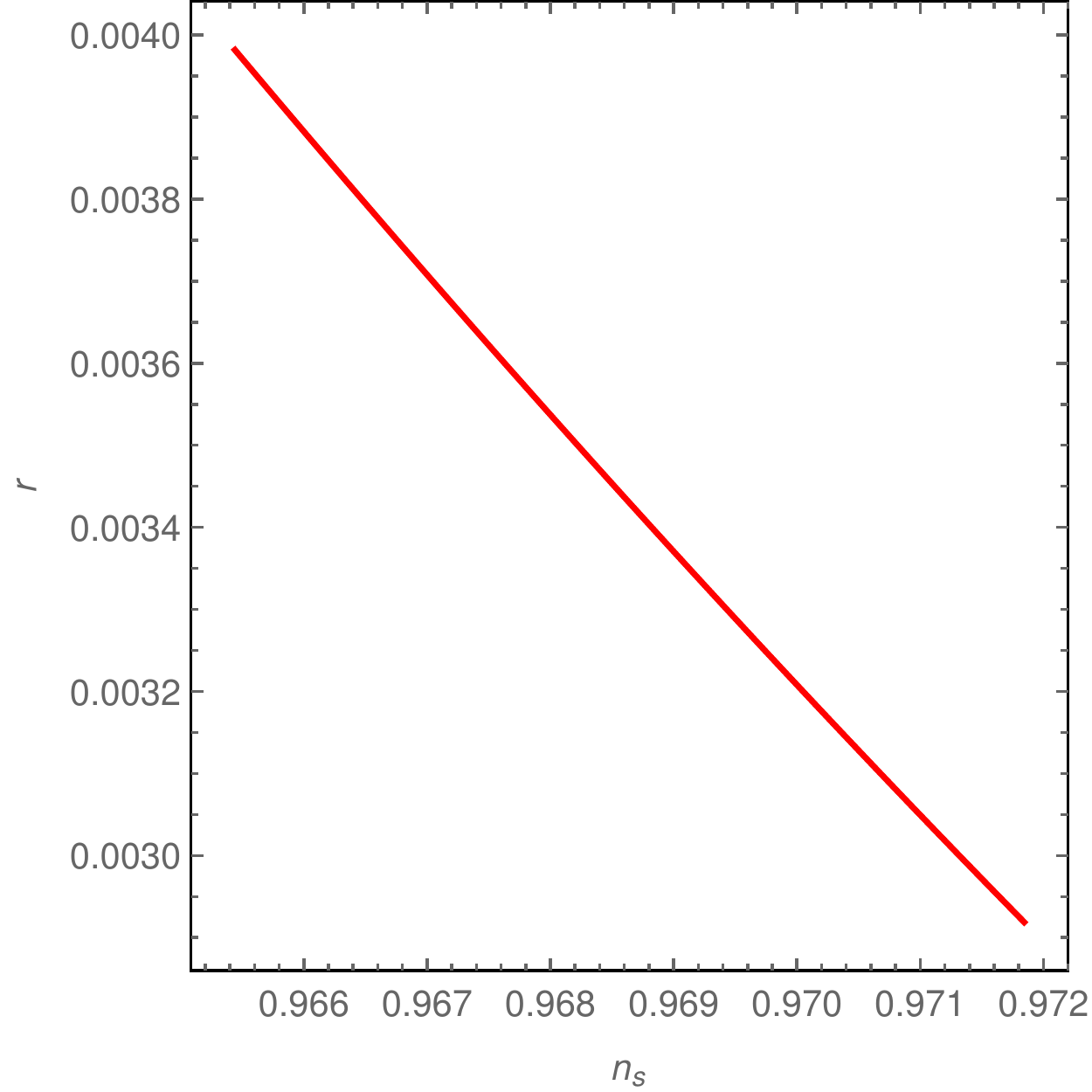}
\includegraphics[width=3.0in,height=2.0in]{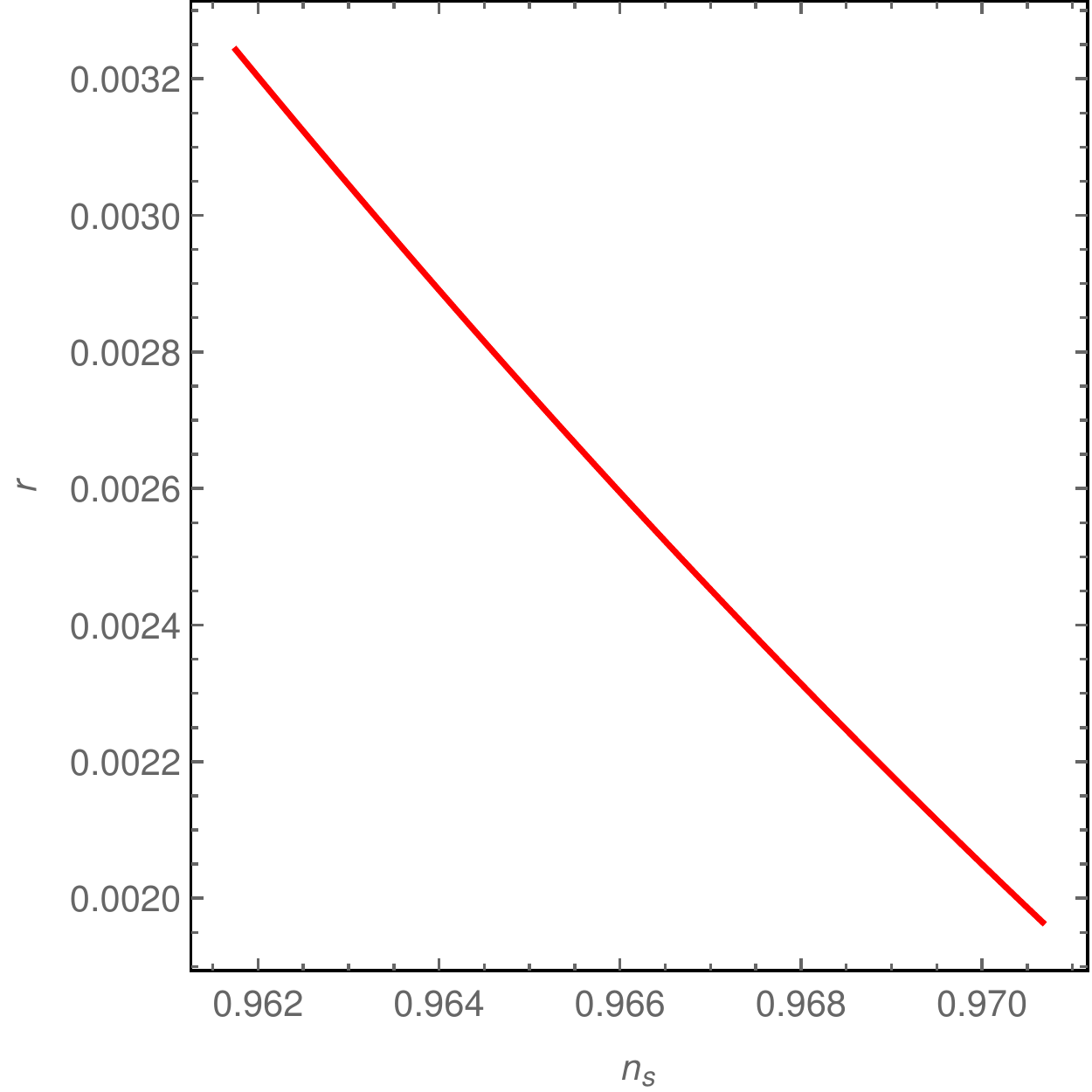}
\caption{{\color{blue}Left Plot}: Parametric plot of $n_s$ vs. $r$ with $\lambda = -0.005$ and $\eta = 1$, in which case, the $n_s$ and $r$ get simultaneously compatible with the Panck data for $N_\mathrm{f} = [50,60]$; {\color{blue}Right Plot}: Parametric plot of $n_s$ vs. $r$ with $\lambda = -0.001$ and $\eta = 1$, which guarantees the consistency of $n_s$ and $r$ provided $N_\mathrm{f} = [50,65]$.}
\label{plot-observable}
\end{center}
\end{figure}

The observable quantities like the spectral tilt for curvature perturbation ($n_s$) and the tensor-to-scalar ratio ($r$) are defined as,
\begin{eqnarray}
 n_s = \frac{\partial\ln{\mathcal{P}_\mathrm{s}}}{\partial\ln{k}}\bigg|_{h.c}~~~~~~~~~~~~~~~~\mathrm{and}~~~~~~~~~~~~~r = \frac{\mathcal{P}_\mathrm{T}}{\mathcal{P}_\mathrm{s}}\bigg|_{h.c}~~,
 \label{ct-1}
\end{eqnarray}
where the suffix 'h.c' indicates the horizon crossing instance of the CMB scale mode on which we are interested to evaluate the observable quantities (the horizon crossing of $k$-th mode occurs at $k = aH$ or $|\tau| = 1/k$). Due to Eq.(\ref{cp-13}) and Eq.(\ref{tp-9}), the $n_s$ and $r$ in the present context comes with the following forms:
\begin{eqnarray}
 n_s = 4 - 2\nu_s~~~~~~~~\mathrm{and}~~~~~~~~r = \frac{8\left(2\epsilon_0 - \delta_0\right)}{\left(1 - \delta_0\right)}\left(\frac{\left(c_\mathrm{s}/2\right)^{2\nu_s}}{\left(c_\mathrm{T}/2\right)^{2\nu_T}}\right)\frac{\Gamma^2(\nu_T)}{\Gamma^2(\nu_s)}~~,
 \label{ct-2}
\end{eqnarray}
where $c_\mathrm{s}$, $c_\mathrm{T}$, $\nu_s$ and $\nu_T$ are obtained above (in terms of slow roll parameters --- see Eq.(\ref{cp-10}), Eq.(\ref{tp-7}), Eq.(\ref{cp-12}) and after Eq.(\ref{tp-8})). Thereby in first order of slow roll parameters, the spectral tilt for curvature perturbation and the tensor-to-scalar ratio from Eq.(\ref{ct-2}) become \cite{Guo:2010jr,Granda:2019jqy},
\begin{eqnarray}
 n_s = \left[1 - 2\epsilon_0 - \frac{2\epsilon_0\epsilon_1 - \delta_0\delta_1}{2\epsilon_0 - \delta_0}\right]_\mathrm{h.c}~~~~~~~~\mathrm{and}~~~~~~~~~
 r = 8\left[2\epsilon_0 - \delta_0\right]_\mathrm{h.c}~~,
 \label{observable-1}
\end{eqnarray}
respectively. The slow roll parameters of Eq.~(\ref{SR-1}) and Eq.~(\ref{SR-2}) immediately lead to $n_s$ and $r$ in terms of model parameters as,
\begin{eqnarray}
 n_s&=&1 - \lambda^2 - \frac{4}{3}V_0\xi_0\eta\left(\lambda + 2\eta\right)\mathrm{e}^{-\left(\lambda + \eta\right)\phi_\mathrm{i}}~~,\nonumber\\
 r&=&8\left(\lambda - \frac{4}{3}V_0\xi_0\eta~\mathrm{e}^{-\left(\lambda + \eta\right)\phi_\mathrm{i}}\right)^2~~,
 \label{observable-2}
\end{eqnarray}
which are calculated at $\phi = \phi_\mathrm{i}$, i.e at the beginning of inflation or equivalently at the horizon crossing of the CMB scale mode. Due to Eq.~(\ref{start phi}), the final forms of $n_s$ and $r$ are give  by.
\begin{eqnarray}
 n_s&=&1 - \lambda^2 - \frac{2\left(\lambda + 2\eta\right)~\mathrm{e}^{\lambda\left(\lambda + \eta\right)N_\mathrm{f}}}
 {\lambda\sqrt{\lambda^2 + 4} - \left(\lambda^2 + 2 - 2\mathrm{e}^{\lambda\left(\lambda + \eta\right)N_\mathrm{f}}\right)}~~,\nonumber\\
 r&=&8\lambda^2\left[\frac{\lambda\sqrt{\lambda^2 + 4} - \left(\lambda^2 + 2\right)}
 {\lambda\sqrt{\lambda^2 + 4} - \left(\lambda^2 + 2 - 2\mathrm{e}^{\lambda\left(\lambda + \eta\right)N_\mathrm{f}}\right)}\right]^2
 \label{observable-3}
\end{eqnarray}
respectively. Having set the stage, we now examine the viability of the model with the recent Planck 2018 data which puts a constraint on $n_s$ and $r$ by,
\begin{eqnarray}
 n_s = 0.9649 \pm 0.0042~~~~~~~~\mathrm{and}~~~~~~~~r < 0.064~~.\nonumber
\end{eqnarray}
It may be noticed from Eq.(\ref{observable-3}) that the theoretical expectations of $n_s$ and $r$ depend on the model parameters $\lambda$ and $\eta$, and on the inflationary e-fold number $N_\mathrm{f}$. The above expressions of $n_s$ and $r$ prove to be simultaneously compatible with the Planck data for suitable values of the model parameters -- these are depicted in the parametric plots of Fig.~[\ref{plot-observable}]. In the left plot of Fig.~[\ref{plot-observable}], we consider $\lambda = -0.005$ and $\eta = 1$, in which case, the theoretical expectations of $n_s$ and $r$ get consistent with the Planck data when the inflationary e-fold lies within $[50,60]$, i.e $N_\mathrm{f} = [50,60]$. Moreover in the right plot of Fig.~[\ref{plot-observable}], the parameters are given by $\lambda = -0.001$ and $\eta = 1$ respectively, which results to the viability of the model provided $N_\mathrm{f}$ lies within $N_\mathrm{f} = [50,65]$. Here it may be mentioned that such ranges of $N_\mathrm{f}$ in both the plots are also consistent in order to resolve the horizon problem. However with $\eta = 1$, $\lambda < -0.007$ seems not to be viable from the perspective of the Planck data and in resolving the horizon problem together.

Furthermore owing to Eq.(\ref{cp-13}), the scalar power spectrum at the horizon exit (i.e at $k|\tau|=1$) comes with the following form,
\begin{eqnarray}
 \mathcal{P}_\mathrm{s} \approx \left(\frac{H_\mathrm{i}^2}{4\pi^2}\right)\frac{1}{\left(2\epsilon_0 - \delta_0\right)}~~,
 \label{p-1}
\end{eqnarray}
which, due to Eq.~(\ref{SR-1}) and Eq.~(\ref{SR-2}), becomes
\begin{eqnarray}
 \mathcal{P}_\mathrm{s} = \left(\frac{1}{12\pi^2}\right)\frac{V_0~e^{-\lambda\phi_\mathrm{i}}}{\left[\lambda - \frac{4}{3}V_0\xi_0\eta\mathrm{e}^{-\left(\lambda + \eta\right)\phi}\right]^2}~~.
 \label{p-2}
\end{eqnarray}
To derive at the above expression, we use the slow roll Eq.(\ref{SR eom}) to determine $H_\mathrm{i}$, in particular, $H_\mathrm{i}^2 = \frac{1}{3}V(\phi_\mathrm{i}) = \frac{V_0}{3}e^{-\lambda\phi_\mathrm{i}}$. According to the Planck observation, the scalar perturbation amplitude is constrained by,
\begin{eqnarray}
 \mathcal{P}_\mathrm{s} = 2.5\times10^{-9}~~.\nonumber
\end{eqnarray}
With $\lambda = -0.005$ and $\eta = 1$, the theoretical expectation of $\mathcal{P}_\mathrm{\psi}$ in Eq.(\ref{p-2}) gets consistent with the observational data if $V_0$ and $\xi_0$ are of the order as,
\begin{eqnarray}
 V_0 \sim 10^{-12} (\mathrm{in~Planck~units})~~~~\mathrm{i.e.}~~~~\kappa^4V_0 = 10^{-12}~~;~~~~~~~\mathrm{and}~~~\kappa^4V_0\xi_0 = 1~~.
 \label{p-3}
\end{eqnarray}
Such constraint on $V_0$ together with the slow roll equation immediately leads to the Hubble parameter at the beginning of inflation (i.e the inflationary energy scale) as,
\begin{eqnarray}
 H_\mathrm{i} \sim 10^{-6} (\mathrm{in~Planck~units})~~.
 \label{p-4}
\end{eqnarray}

Thus as a whole, the scalar-Einstein-GB theory with $V_\mathrm{I}(\phi) = V_0\mathrm{e}^{-\lambda\phi}$ and $\xi_\mathrm{I}(\phi) = \xi_0\mathrm{e}^{\eta\phi}$, triggers a viable inflationary scenario during the early universe. In particular -- (1) the inflation has an exit at a finite e-fold, which is indeed consistent with the resolution of horizon and flatness problems, (2) the scalar and tensor perturbations prove to be ghost free and do not suffer from gradient instability, (3) the scalar spectral index for curvature perturbation and the tensor-to-scalar ratio turn out to be simultaneously compatible with the recent Planck data for suitable values of the parameters as shown in Fig.~[\ref{plot-observable}], and (4) the consistency of the scalar perturbation amplitude with the Planck data demands the inflationary energy scale to be of the order $\sim 10^{-6}$ (in Planck units).

\section{From inflation to reheating}\label{sec-reh}

After the inflation ends the universe enters to reheating era, during which, the scalar field potential and the GB coupling function in the present context follow Eq.(\ref{reh-pot}). During the reheating era, the scalar field energy density decays to radiation energy by a certain decay width considered generally to be a constant and is represented by $\Gamma$. Here we would like to mention that in the current scenario, the scalar field energy density is contributed from its canonical part as well as from the GB coupling with the scalar field, i.e we may write,
\begin{eqnarray}
 \rho_\mathrm{\phi} = \frac{1}{2}\dot{\phi}^2 + V(\phi) - 12\dot{\xi}H^3~~,
 \label{reh-1}
\end{eqnarray}
and moreover, the effective pressure of the scalar field is given by,
\begin{eqnarray}
 p_\mathrm{\phi} = \dot{\phi}^2 + 4\ddot{\xi}H^2 + 4\dot{\xi}H\left(2\dot{H} - H^2\right) - \rho_\mathrm{\phi}~~.
 \label{reh-2}
\end{eqnarray}
Consequently the effective equation of state (EoS) parameter during the reheating era comes as
\begin{eqnarray}
 w_\mathrm{eff} = \frac{p_\mathrm{\phi}}{\rho_\mathrm{\phi}} = -1 + \frac{\dot{\phi}^2 + 4\ddot{\xi}H^2 + 4\dot{\xi}H\left(2\dot{H} - H^2\right)}
 {\frac{1}{2}\dot{\phi}^2 + V(\phi) - 12\dot{\xi}H^3}~~.
 \label{reh-3}
\end{eqnarray}
Clearly the presence of the GB term affects the reheating era by two ways: (a) by implicit way through the dynamics of the scalar field, and (b) by explicit way through the derivatives of $\xi(\phi)$ present in the above expression of $w_\mathrm{eff}$. After the end of inflation, the Hubble parameter is much larger than the decay width (i.e the condition $H \gg \Gamma$ holds), owing to which, the comoving scalar field energy density remains conserved with the cosmological expansion of the universe. However the Hubble parameter continuously decreases and eventually gets comparable to $\Gamma$ when the scalar field effectively decays to radiation and indicates the end of reheating. Thus as a whole, the comoving energy density of the scalar field remains conserved (or equivalently, the decaying of the scalar field is negligible with respect to the Hubble expansion) during $H \gg \Gamma$, and finally, the scalar field instantaneously decays to radiation at the end of reheating when $H = \Gamma$ satisfies. As a result, the Hubble parameter follows a power law solution during the reheating era, in particular, we take the following power law ansatz :
\begin{eqnarray}
 H(t) = m/t~~,
 \label{reh-4}
\end{eqnarray}
during the reheating era, where $m$ is the corresponding exponent. The above solution of $H(t)$ actually generates from the following scalar field potential and the GB coupling function:
\begin{eqnarray}
 V_\mathrm{R}(\phi) = V_1\mathrm{e}^{2\left(\phi-\phi_\mathrm{s}\right)/\phi_\mathrm{0}}
 ~~~~\mathrm{and}~~~~\xi_\mathrm{R}(\phi) = \xi_1\mathrm{e}^{-2\left(\phi - \phi_\mathrm{s}\right)/\phi_\mathrm{0}}~,
\end{eqnarray}
that are operating during the reheating stage (also demonstrated earlier in Eq.(\ref{reh-pot})). Here it may be mentioned that the forms of $V_\mathrm{R}(\phi)$ (and $\xi_\mathrm{R}(\phi)$) are different that that of operating during inflation (recall the scalar potential and the GB coupling function during inflation are symbolized by $V_\mathrm{I}(\phi)$ and $\xi_\mathrm{I}(\phi)$ respectively, see Eq.(\ref{inf-pot})). The respective forms of $V(\phi)$ and $\xi(\phi)$ are joined smoothly at the junction between inflation-to-reheating for suitable values of parameters,
and their full forms that describe from the inflation to the reheating stage of the universe will be demonstrated in Sec.~[\ref{sec-complete}]. Below, we will present the scalar field solution during the reheating era corresponds to $V_\mathrm{R}(\phi)$ and $\xi_\mathrm{R}(\phi)$ by considering the above mentioned solution of the Hubble parameter. Eq.(\ref{reh-4}) immediately leads to the reheating EoS parameter as,
\begin{eqnarray}
 w_\mathrm{eff} = -1 - \frac{2\dot{H}}{3H^2} = -1 + \frac{2}{3m}
 \label{reh-5}
\end{eqnarray}
which is actually a constant, as expected. In the above equation, we use Eq.(\ref{reh-3}) along with the background field equations as $3H^2 = \rho_\mathrm{\phi}$ and $2\dot{H} + 3H^2 = -p_\mathrm{\phi}$, where $\rho_\mathrm{\phi}$ and $p_\mathrm{\phi}$ are shown in Eq.(\ref{reh-1}) and Eq.(\ref{reh-2}) respectively. Being $w_\mathrm{eff}$ is a constant, the scalar field energy density varies as $\rho_\mathrm{\phi} \propto a^{-3\left(1 + w_\mathrm{eff}\right)}$ with the scale factor of the universe, or equivalently the comoving scalar energy density, i.e $\rho_\mathrm{\phi}\times a^{3\left(1 + w_\mathrm{eff}\right)}$, proves to be conserved.\\

With $H(t) = \frac{m}{t}$ along with the $V_\mathrm{R}(\phi) = V_1\mathrm{e}^{2\left(\phi-\phi_\mathrm{s}\right)/\phi_\mathrm{0}}$ and the $\xi_\mathrm{R}(\phi) = \xi_1\mathrm{e}^{-2\left(\phi-\phi_\mathrm{s}\right)/\phi_\mathrm{0}}$ (i.e the scalar potential and the GB coupling function during the reheating era) shown in Eq.(\ref{reh-pot}), the field equations (see Eq.(\ref{FRW-1}), Eq.(\ref{FRW-2}) and Eq.(\ref{scalar eom})) immediately lead to the scalar field's evolution during the reheating phase as,
\begin{eqnarray}
 \phi(t) = \phi_\mathrm{s} - \phi_\mathrm{0}~\mathrm{ln}{t}~~,
 \label{reh-6}
\end{eqnarray}
provided $\phi_\mathrm{0}$ is connected with $V_\mathrm{1}$ and $\xi_\mathrm{1}$ by the following way:
\begin{eqnarray}
 V_\mathrm{1}&=&\frac{\left(5m^2 - m\right)}{(1 + m)}\left(\frac{\phi_\mathrm{0}^2}{2m} - 1\right) + 3m^2 - m~~,\nonumber\\
 \xi_\mathrm{1}&=&\frac{1}{4m(1 + m)}\left(\frac{\phi_\mathrm{0}^2}{2m} - 1\right)~~.
 \label{reh-7}
\end{eqnarray}
For having simpler expressions, we denote $\frac{1}{(1 + m)}\left(\frac{\phi_\mathrm{0}^2}{2m} - 1\right) = c_1$ (say) from onwards. Thus during the reheating era described by $V_\mathrm{R}(\phi) = V_1\mathrm{e}^{2\left(\phi-\phi_\mathrm{s}\right)/\phi_\mathrm{0}}$ and $\xi_\mathrm{R}(\phi) = \xi_1\mathrm{e}^{-2\left(\phi-\phi_\mathrm{s}\right)/\phi_\mathrm{0}}$, the Hubble parameter and the scalar field follow Eq.(\ref{reh-4}) and Eq.(\ref{reh-6}) respectively where $\phi_\mathrm{0}$ being connected with $V_\mathrm{1}$ and $\xi_\mathrm{1}$ by Eq.(\ref{reh-7}). Having obtained the solutions of $H(t)$ and $\phi(t)$, it is worthwhile to examine whether the background evolutions are stable attractor or not -- this is the subject of the next subsection.

\subsection*{Dynamical analysis of the reheating dynamics}
In the context of scalar-Einstein-GB theory of gravity where the cosmological field equations are given by Eq.(\ref{FRW-1}), Eq.(\ref{FRW-2}) and Eq.(\ref{scalar eom}) respectively, the dynamical variables may be taken as follows:
\begin{eqnarray}
 x = \dot{\phi}^2/\left(6H^2\right)~~~~~~~~,~~~~~~~~y = V(\phi)/\left(3H^2\right)~~~~~~~~\mathrm{and}~~~~~~~z = -4\dot{\xi}H~~.
 \label{dynamical-1}
\end{eqnarray}
With these variables, Eq.(\ref{FRW-1}) becomes,
\begin{eqnarray}
 x + y + z = 1~~.
 \label{dyn-2}
\end{eqnarray}
For the purpose of dynamical analysis, it will be useful if we transform the cosmic time to e-fold number ($N$) by the prescription $\frac{d}{dt} = H\frac{d}{dN}$. In terms of the dynamical variables ($x(N)$, $y(N)$ and $z(N)$), the governing field equations in the present context take the following form:
\begin{eqnarray}
 \frac{dx}{dN}&=&-\left(6 + \frac{2}{H}\frac{dH}{dN}\right)x - \frac{dy}{dN} - \left(\frac{2}{H}\frac{dH}{dN}\right)y - z\left(1 + \frac{1}{H}\frac{dH}{dN}\right)~~,\nonumber\\
 \frac{dy}{dN}&=&\left[-\frac{2}{H}\frac{dH}{dN} + \frac{1}{V(\phi)}\frac{dV(\phi)}{dN}\right]y~~,\nonumber\\
 \frac{dz}{dN}&=&\left[\frac{2}{H}\frac{dH}{dN} + \frac{1}{\xi'(\phi)}\frac{d\xi'(\phi)}{dN} + \frac{1}{2x}\frac{dx}{dN}\right]z~~.
 \label{dyn-3}
\end{eqnarray}
At the background dynamics during the reheating era, in particular for $H(t) = \frac{m}{t}$ and $\phi(t) = \phi_\mathrm{s} - \phi_\mathrm{0}~\mathrm{ln}{t}$, the dynamical variables become constant and are given by,
\begin{eqnarray}
 x_\mathrm{bg}&=&\frac{1}{3}\left(c_1 + \frac{c_1}{m} + \frac{1}{m}\right)~~,\nonumber\\
 y_\mathrm{bg}&=&\frac{1}{3}\left(5c_1 - \frac{c_1}{m} + 3 - \frac{1}{m}\right)~~,\nonumber\\
 z_\mathrm{bg}&=&-2c_1~~,
 \label{dyn-4}
\end{eqnarray}
respectively, where recall that $c_1 = \frac{1}{(1 + m)}\left(\frac{\phi_\mathrm{0}^2}{2m} - 1\right)$. Using the dynamical Eq.(\ref{dyn-3}), one can show that
\begin{eqnarray}
 \frac{dx}{dN}\bigg|_\mathrm{bg} = \frac{dy}{dN}\bigg|_\mathrm{bg} = \frac{dz}{dN}\bigg|_\mathrm{bg} = 0~~,
 \label{dyn-5}
\end{eqnarray}
i.e the background dynamics obtained in Eq.(\ref{dyn-4}) is indeed a fixed point in the ($x$, $y$, $z$) phase space. In order to examine whether the fixed point ($x_\mathrm{bg}$, $y_\mathrm{bg}$, $z_\mathrm{bg}$) is a stable attractor, we consider a linear perturbation of the dynamical variables around the fixed point. It turns out that the perturbation described by $u(N) = x - x_\mathrm{bg}$ evolves as:
\begin{eqnarray}
 u(N) \sim \mathrm{exp}\left(-\gamma N\right)~~,
 \label{dyn-6}
\end{eqnarray}
where the exponent $\gamma$ is given by,
\begin{eqnarray}
 \gamma = \frac{\left(3c_1 + \frac{7c_1}{m} + \frac{6}{m} - \frac{2c_1}{m^2 - \frac{2}{m^2}}\right)}
 {\left(c_1 + \frac{c_1}{m} + \frac{1}{m}\right)}~~.
 \label{dyn-7}
\end{eqnarray}
Clearly for $\gamma > 0$, the perturbation decays and hence the background dynamics prove to be a stable attractor. Eq.(\ref{dyn-7}) indicates that the condition $\gamma > 0$ holds for
\begin{eqnarray}
 c_1 > \left(\frac{2-6m}{3m^2 + 7m -2}\right)~~~~~~~\mathrm{and}~~~~~~~~m > \frac{\sqrt{73} - 7}{6}~~.
 \label{dyn-8}
\end{eqnarray}
Moreover $m$ should be less than unity in order to have a deceleration expansion of the universe during the reheating stage (recall the Hubble parameter during the reheating follows $H = m/t$, see Eq.(\ref{reh-4})). Combining $m < 1$ with Eq.(\ref{dyn-8}), we get,
\begin{eqnarray}
 c_1 > \left(\frac{2-6m}{3m^2 + 7m -2}\right)~~~~~~~\mathrm{and}~~~~~~~~\frac{\sqrt{73} - 7}{6} < m < 1
 \label{dyn-9}
\end{eqnarray}
which ensure the stability of the reheating dynamics governed by Eq.(\ref{reh-4}) and Eq.(\ref{reh-6}) (or equivalently, Eq.(\ref{dyn-4})) respectively. The above two conditions in Eq.(\ref{dyn-9}) immediately lead to $V_1 = 5c_1m^2 - c_1m + 3m^2 - m > 0$ which in turn makes the $V_\mathrm{R}(\phi)$ positive definite. Here we would like to mention that a positive definite potential also hints to stable scenario, actually a scalar potential which is unbounded from below results to an unstable situation. Eq.(\ref{reh-5}) depicts that due to the stability condition on $m$, the effective EoS parameter during the reheating stage lies within the following range:
\begin{eqnarray}
 -\frac{1}{3} < w_\mathrm{eff} < 1.56~~.
 \label{dyn-10}
\end{eqnarray}
Therefore a stable dynamics in the context of scalar-Einstein-GB gravity allows the effective EoS parameter during the reheating stage to go beyond unity and reach up-to $\mathrm{max}\left(w_\mathrm{eff}\right) = 1.56$. However in canonical scalar-tensor (CST) theory where the GB coupling is absent, the reheating EoS parameter can have the maximum value of unity. Actually $w_\mathrm{eff} > 1$ in scalar-tensor theory makes the background dynamics unstable -- this can be demonstrated as follows. In absence of the GB coupling, $\xi_\mathrm{R}(\phi) = 0$ or or equivalently $c_1 = 0$ (see Eq.(\ref{reh-7})), and thus the scalar potential during the reheating stage takes the form as (see Eq,(\ref{reh-7}) and Eq.(\ref{reh-pot})):
\begin{eqnarray}
 V_\mathrm{R}(\phi) = \left(3m^2 - m\right)e^{2\left(\phi - \phi_\mathrm{s}\right)/\phi_\mathrm{0}}~~.
 \label{dyn-11}
\end{eqnarray}
Moreover the dynamical variables in CST theory are given by,
\begin{eqnarray}
 x = \dot{\phi}^2/\left(6H^2\right)~~~~~~~\mathrm{and}~~~~~~~y = |V(\phi)|/\left(3H^2\right)~~.
 \label{dyn-12}
\end{eqnarray}
In the CST theory, $\xi(\phi) = 0$ that leads to $z = 0$, and thus the dynamical phase space ($x$, $y$) becomes two dimensional. Consequently the background reheating dynamics for the CST theory described by $V_\mathrm{R}(\phi) = \left(3m^2 - m\right)e^{2\left(\phi - \phi_\mathrm{s}\right)/\phi_\mathrm{0}}$ is represented by a fixed point in ($x$, $y$) phase space, and given by,
\begin{eqnarray}
 x_\mathrm{bg} = \frac{1}{3m}~~~~~~~~~~\mathrm{and}~~~~~~~~~~~y_\mathrm{bg} = 1 - \frac{1}{3m}~~,
 \label{dyn-13}
\end{eqnarray}
respectively. Performing the linear perturbation analysis around the fixed point ($x_\mathrm{bg}$, $y_\mathrm{bg}$),
we find that the perturbation variable $u(N) = x - x_\mathrm{bg}$ goes as $u(N) \sim e^{-\gamma N}$ with $\gamma = 2\left(3 - \frac{1}{m}\right)$ -- this is consistent with Eq.(\ref{dyn-7}) for $c_1 = 0$. Therefore in the CST theory, the perturbation decays for $m > \frac{1}{3}$, while it grows for $m < \frac{1}{3}$. On other hand, $m < \frac{1}{3}$ leads to $w_\mathrm{eff} = -1 + \frac{2}{3m} > 1$ and vice-versa. This clearly demonstrates that in the canonical scalar-tensor theory, the
stable condition of reheating dynamics demands $w_\mathrm{eff} < 1$. Actually in the CST theory, the condition $w_\mathrm{eff} > 1$ makes the $V_\mathrm{R}(\phi)$ unbounded from below (see Eq.(\ref{dyn-11})), which in turn makes the background dynamics unstable. On contrary, in the context of scalar-Einsten-GB gravity, the presence of GB coupling function ensures the stability of the background reheating dynamics even for $w_\mathrm{eff}> 1$ (in particular for $w_\mathrm{eff} < 1.56$).

This may have important consequences on primordial gravitational waves' (PGWs) evolution, as the reheating EoS parameter is expected to play a significant role on enhancing the PGW's amplitude \cite{Haque:2021dha}. One of our authors showed in \cite{Haque:2021dha} that in the context of scalar-tensor theory, $w_\mathrm{eff} > \frac{1}{3}$ results to an enhancement on PGW's amplitude observed today, and the corresponding spectral tilt is proportional to $\propto \left(3w_\mathrm{eff} - 1\right)$ -- thus larger $w_\mathrm{eff}$ leads to more enhanced PGWs' spectrum today. This, along with the fact that $w_\mathrm{eff}$ in the scalar-Einstein-GB theory reaches up to $\mathrm{max}(w_\mathrm{eff}) = 1.56$ which is larger than the $\mathrm{max}(w_\mathrm{eff})$ of scalar-tensor theory, hints our expectation that the GB theory of gravity results to more enhancement on PGWs' amplitude compared to that of in scalar-tensor theory. However this needs proper investigation and we expect to study it in near future.

\subsection*{Reheating e-folding and reheating temperature}
As mentioned earlier, the Hubble parameter during the reheating era goes as $H = \frac{m}{t}$ or equivalently $H \propto a^{-\frac{3}{2}\left(1 + w_\mathrm{eff}\right)}$
(where $w_\mathrm{eff} = -1 + \frac{2}{3m}$) in terms of the scale factor of the universe. Thus we may write the Hubble parameter as,
\begin{eqnarray}
 H(a) = H_\mathrm{f}\left(\frac{a}{a_\mathrm{f}}\right)^{-\frac{3}{2}\left(1 + w_\mathrm{eff}\right)}~~,
 \label{et-1}
\end{eqnarray}
where the suffix 'f' with a quantity refers to the quantity at the end of inflation. In this note, we also mention that a suffix 're' denotes the end of reheating. Recall that th end of reheating is depicted when the Hubble parameter becomes comparable to the decay width of the scalar field, i.e when $H = \Gamma$ satisfies. As a result, Eq.(\ref{et-1}) may express the $\Gamma$ in terms of $H_\mathrm{f}$, $w_\mathrm{eff}$ and the e-fold number of the reheating era (defined by $N_\mathrm{re} = \ln{\left(a_\mathrm{re}/a_\mathrm{f}\right)}$) as,
\begin{eqnarray}
 \Gamma = H_\mathrm{f}~\mathrm{exp}\left[-\frac{3}{2}N_\mathrm{re}\left(1 + w_\mathrm{eff}\right)\right]~~.
 \label{et-2}
\end{eqnarray}
Moreover, the $\Gamma$ can also be expressed by $\Gamma^2 = \frac{1}{3}\rho_\mathrm{re}$ (in the unit of $\kappa = 1$) from the FRW equation, where $\rho_\mathrm{re}$ represents the energy density at the end of reheating when the temperature of the universe is generally symbolized by $T_\mathrm{re}$ (also known as the reheating temperature). Therefore, we may write
\begin{eqnarray}
N_\mathrm{re}&=&\frac{1}{3\left(1 + w_\mathrm{eff}\right)}\ln{\left(\frac{3H_\mathrm{f}^2}{\rho_\mathrm{re}}\right)}~~,\label{et-3a}\\
 \rho_\mathrm{re}&=&\left(\frac{\pi^2g_\mathrm{re}}{30}\right)T_\mathrm{re}^4~~,
 \label{et-3}
\end{eqnarray}
where $g_\mathrm{re}$ denotes the relativistic degrees of freedom. Since the whole energy density of the scalar field is considered to decay at the end of the reheating, we may argue that the entropy of the universe remains conserved from the end of reheating to the present epoch. This connects the reheating temperature ($T_\mathrm{re}$) to the present temperature ($T_0$) of the universe via \cite{Cook:2015vqa},
\begin{eqnarray}
 T_\mathrm{re} = T_0\left(\frac{a_0}{a_\mathrm{re}}\right)\left(\frac{43}{11g_\mathrm{re}}\right)^{1/3}~~,
 \label{N1}
\end{eqnarray}
where $a_0$ and $a_\mathrm{re}$ are the scale factor of the universe at the present epoch and at the end of reheating respectively and and $T_0 = 2.93\mathrm{K}$. Here we would like to mention that $T_\mathrm{re}$ is defined as the temperature of the universe at the end of reheating, and thus, the presence of only $\{T_\mathrm{re},T_0\}$ and $\{a_\mathrm{re},a_0\}$ (along with the relativistic degrees of freedom) in the above equation clearly indicates that Eq.(\ref{N1}) uses the informations only from the end of reheating to the present epoch. In this sense, Eq.(\ref{N1}) is general that it does not assume any particular inflation or reheating model, and thereby it is equally valid in the present context of scalar-Einstein-GB gravity theory. Moreover $\frac{a_0}{a_\mathrm{re}}$ can be equivalently expressed as,
\begin{eqnarray}
 \frac{a_0}{a_\mathrm{re}} = \left(\frac{a_0H_\mathrm{i}}{k}\right)\left(\frac{a_\mathrm{i}}{a_\mathrm{re}}\right) = \left(\frac{a_0H_\mathrm{i}}{k}\right)e^{-\left(N_\mathrm{f} + N_\mathrm{re}\right)}~~.
 \label{N2}
\end{eqnarray}
Here $a_\mathrm{i}$ is the scale factor at the beginning of inflation, or equivalently, when the CMB scale ($k$) crosses the horizon at $k = a_\mathrm{i}H_\mathrm{i}$ (with $\frac{k}{a_0} = 0.05\mathrm{Mpc}^{-1}$), and $H_\mathrm{i} \sim 10^{13}\mathrm{GeV}$ represents the inflationary energy scale (or $H_\mathrm{i} \sim 10^{-6}$ in Planck units, see Eq.(\ref{p-4})). To derive Eq.(\ref{N2}), one needs to remember the following two points: (a) the horizon crossing condition of the CMB scale, i.e $k = a_\mathrm{i}H_\mathrm{i}$, and (b) the definition of e-folding number, in particular $\frac{a_\mathrm{i}}{a_\mathrm{re}} = e^{-\left(N_\mathrm{f} + N_\mathrm{re}\right)}$. This indicates that similar to the previous equation, Eq.(\ref{N2}) is also valid for any inflation or reheating model under consideration, and hence, we can safely use Eq.(\ref{N2}) in the present context. Plugging back the above expression of $\frac{a_0}{a_\mathrm{re}}$ to Eq.(\ref{N1}) yields the reheating temperature as,
\begin{eqnarray}
 T_\mathrm{re} = H_\mathrm{i}\left(\frac{43}{11g_\mathrm{re}}\right)^{1/3}\left(\frac{T_0}{k/a_0}\right)e^{-\left(N_\mathrm{f} + N_\mathrm{re}\right)}~~,
 \label{et-4}
\end{eqnarray}
in the current scalar-Einstein-GB model. As we have demonstrated above that Eq.(\ref{et-4}) is valid for all inflation (or reheating) model, however the informations of the specific inflation (or reheating) model under consideration will be entered to the reheating temperature through $H_\mathrm{i}$, $N_\mathrm{f}$ and $N_\mathrm{re}$ present in Eq.(\ref{et-4}). Below we will show how the $T_\mathrm{re}$ gets affected and put constraints on the parameters of the current model. By using Eq.(\ref{et-3}) and Eq.(\ref{et-4}), we further get $\rho_\mathrm{re}$ in terms of the inflationary parameters and the reheating e-fold number as,
\begin{eqnarray}
 \rho_\mathrm{re} = H_\mathrm{i}^4\left(\frac{\pi^2g_\mathrm{re}}{30}\right)\left(\frac{43}{11g_\mathrm{re}}\right)^{4/3}\left(\frac{T_0}{k/a_0}\right)^4e^{-4\left(N_\mathrm{f} + N_\mathrm{re}\right)}~~,
 \label{et-5}
\end{eqnarray}
Plugging back the above expression of $\rho_\mathrm{re}$ into Eq.(\ref{et-3a}) along with a little bit of simplification yield the final expression of $N_\mathrm{re}$ as follows \cite{Cook:2015vqa}:
\begin{eqnarray}
 N_\mathrm{re} = \frac{4}{\left(1 - 3w_\mathrm{eff}\right)}\left\{-\frac{1}{4}\ln{\left(\frac{30}{\pi^2g_\mathrm{re}}\right)} - \frac{1}{3}\ln{\left(\frac{11g_\mathrm{re}}{43}\right)} - \ln{\left(\frac{k/a_0}{T_0}\right)} - \ln{\left[\frac{\left(3H_\mathrm{f}^2\right)^{1/4}}{H_\mathrm{i}}\right]} - N_\mathrm{f}\right\}~~.
 \label{et-6}
\end{eqnarray}
Clearly the $N_\mathrm{re}$ is represented in terms of the inflationary parameters (like $N_\mathrm{f}$, $H_\mathrm{I}$) and the reheating EoS parameter ($w_\mathrm{eff}$). With $\frac{k}{a_0} = 0.05\mathrm{Mpc}^{-1} \approx 10^{-40}\mathrm{GeV}$ (where the conversion $1\mathrm{Mpc}^{-1} = 10^{-38}\mathrm{GeV}$ may be useful), $T_0 = 2.93K$ and $g_\mathrm{re} = 100$, one simply gets,
\begin{eqnarray}
 N_\mathrm{re} = \frac{4}{\left(1 - 3w_\mathrm{eff}\right)}\left\{61.6 - \ln{\left[\frac{\left(3H_\mathrm{f}^2\right)^{1/4}}{H_\mathrm{i}}\right]} - N_\mathrm{f}\right\}~~.
 \label{et-7}
\end{eqnarray}
The quantity $H_\mathrm{i}$, i.e the Hubble parameter at the beginning of inflation, can be determined from slow roll Eq.(\ref{SR eom}), in particular,
\begin{eqnarray}
 H_\mathrm{i}^2 = \left(\frac{V_0}{3}\right)e^{-\lambda\phi_\mathrm{i}}~~,
 \label{et-8}
\end{eqnarray}
where $\phi_\mathrm{i}$ is the value of the scalar field at the beginning of inflation when the CMB mode $k = 0.05\mathrm{Mpc}^{-1}$ (on which we are interested) crosses the horizon) and shown in Eq.(\ref{start phi}). On the other hand the end of inflation is designated by $\Delta_0 = -1$ (see the discussion after Eq.(\ref{SR-2})), and thus Eq.(\ref{FRW-1}) provides the Hubble parameter at the end of inflation as,
\begin{eqnarray}
 H_\mathrm{f}^2 = \left(\frac{V_0}{6}\right)e^{-\lambda\phi_\mathrm{f}}~~,
 \label{et-9}
\end{eqnarray}
with $\phi_\mathrm{f}$ being obtained earlier in Eq.(\ref{end phi}). Due to these expressions of $H_\mathrm{i}$ and $H_\mathrm{f}$, Eq.(\ref{et-7}) provides $N_\mathrm{re}$ in terms of the model parameters in the present context as,
\begin{eqnarray}
 N_\mathrm{re} = \frac{4}{\left(1 - 3w_\mathrm{eff}\right)}\left\{61.6 - \frac{1}{4}\ln{\left(\frac{9e^{\lambda\left(2\phi_\mathrm{i} - \phi_\mathrm{f}\right)}}{2V_0}\right)} - N_\mathrm{f}\right\}~~.
 \label{et-10}
\end{eqnarray}
Consequently the reheating temperature from Eq.(\ref{et-4}) becomes,
\begin{eqnarray}
 T_\mathrm{re} = \left(\frac{V_0e^{-\lambda\phi_\mathrm{i}}}{3}\right)^{1/2}\left(\frac{43}{11g_\mathrm{re}}\right)^{1/3}\left(\frac{T_0}{k/a_0}\right)
 \mathrm{exp}\left[\frac{3\left(1 + w_\mathrm{eff}\right)}{\left(1 - 3w_\mathrm{eff}\right)}N_\mathrm{f} - \frac{4}{\left(1 - 3w_\mathrm{eff}\right)}\left\{61.6 - \frac{1}{4}\ln{\left(\frac{9e^{\lambda\left(2\phi_\mathrm{i} - \phi_\mathrm{f}\right)}}{2V_0}\right)}\right\}\right]~~.
 \label{et-11}
\end{eqnarray}
Recall from Eq.(\ref{p-3}) that $V_0$ is of the order $\sim 10^{-12}$ (in Planck units) in order to have a compatible scalar perturbation amplitude with respect to the Planck data. Regarding the inflationary phenomenology, we have considered two different sets of model parameters: (1) $\lambda = -0.005$, $\eta = 1$; and (2) $\lambda = -0.001$, $\eta = 1$ respectively. For these two sets, we have found that the theoretical expectations of $n_s$ and $r$ get consistent with the observational data provided the inflationary e-fold number lies within $N_\mathrm{f} = [50,60]$ for the first set and $N_\mathrm{f} = [50,65]$ for the second set. We now examine whether such constraints on the model parameters coming from the inflationary phenomenology are supported or ruled out or get further constrained by the reheating stage. For this purpose, we give the plots for $N_\mathrm{re}$ vs. $N_\mathrm{f}$ and $T_\mathrm{re}$ vs. $N_\mathrm{f}$ for various values of $w_\mathrm{eff}$ (here we need to keep in mind that $w_\mathrm{eff}$ ranges within $-\frac{1}{3} < w_\mathrm{eff} < 1.56$ from the stability of the reheating dynamics, see Eq.(\ref{dyn-10})). In particular, we take $w_\mathrm{eff} = 0, \frac{1}{6}, \frac{2}{3}, 1, 1.56$ respectively.

 \begin{figure}[!h]
\begin{center}
\centering
\includegraphics[width=3.0in,height=2.0in]{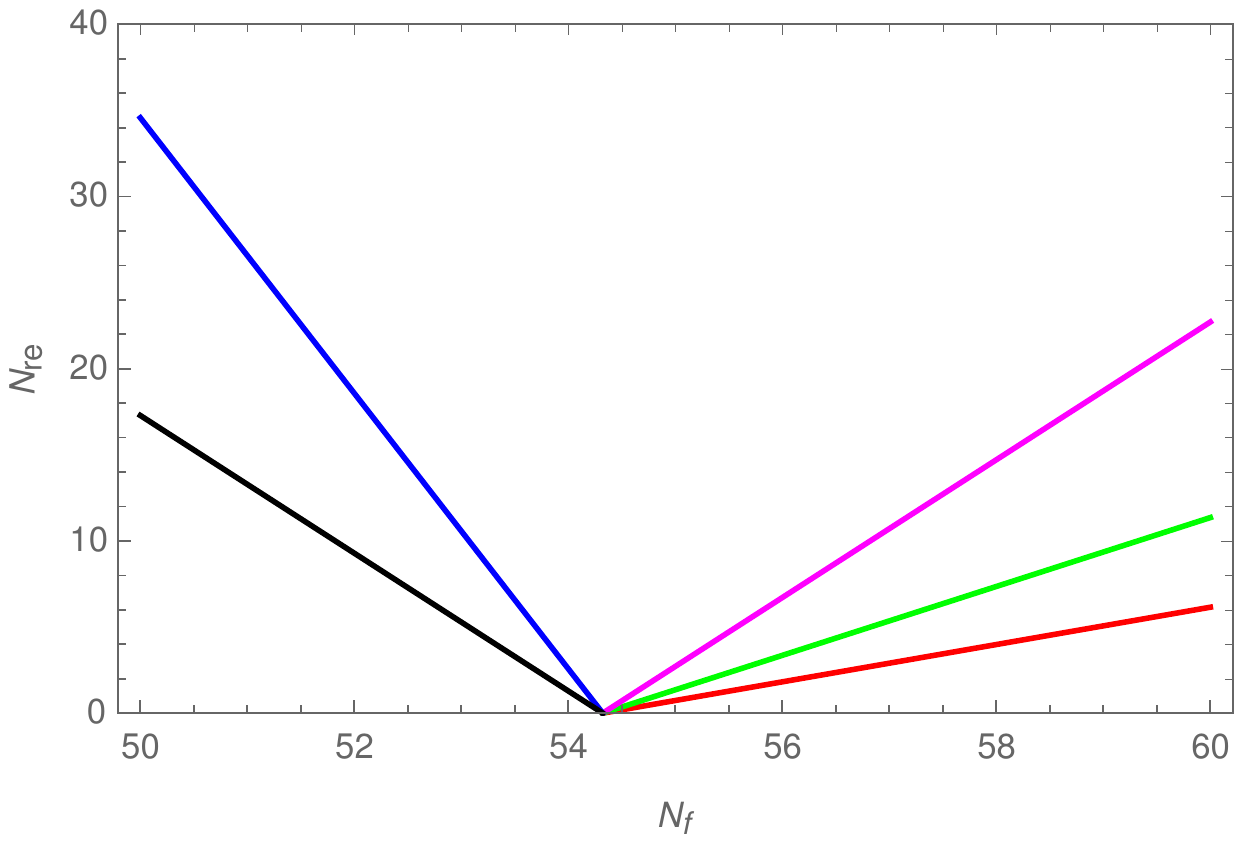}
\includegraphics[width=3.0in,height=2.0in]{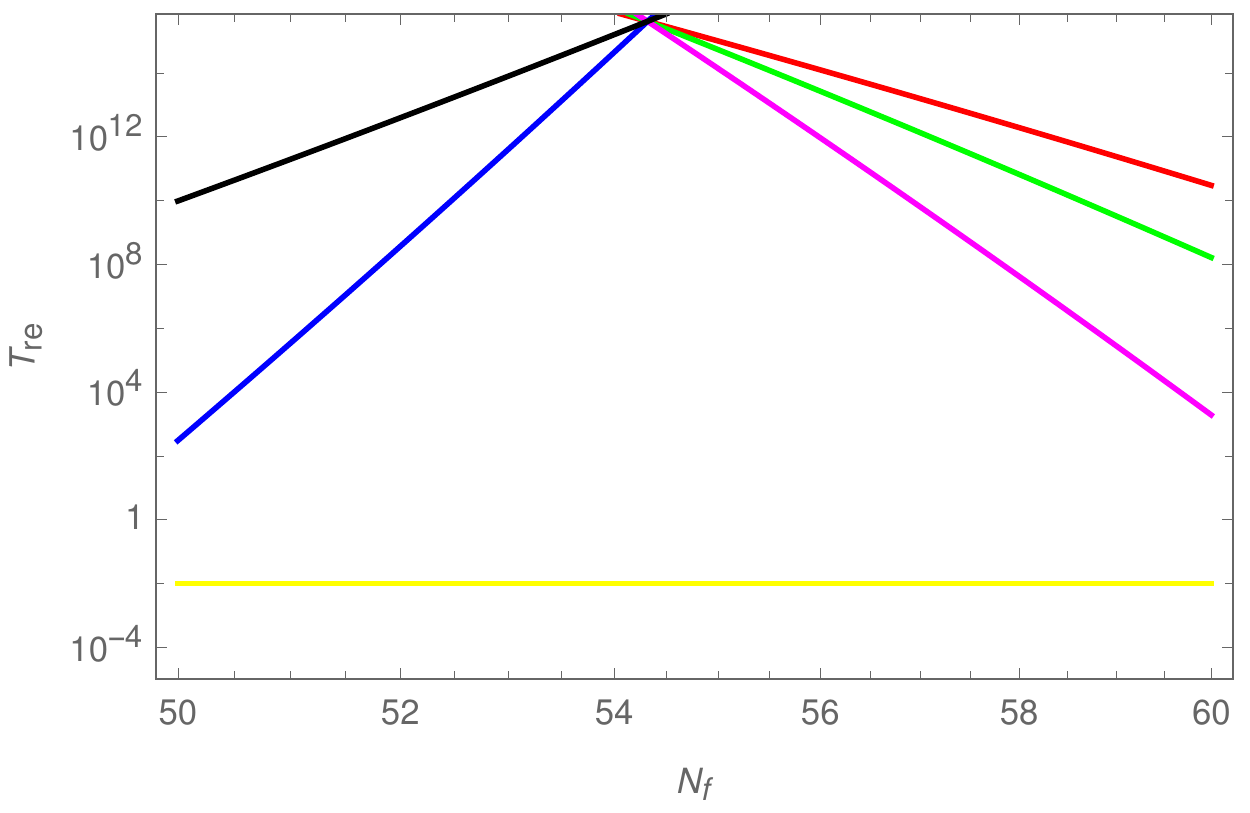}
\caption{{\color{blue}Left Plot}: $N_\mathrm{re}$ vs. $N_\mathrm{f}$; {\color{blue}Right Plot}: $T_\mathrm{re}$ (in GeV unit) vs. $N_\mathrm{f}$ for various values of $w_\mathrm{eff}$. In both the plots, we consider $\lambda = -0.005$ $\eta = 1$ and $V_0 = 10^{-12}$ (in Planck units). The reheating EoS parameter is taken as $w_\mathrm{eff} = 0 ~(\mathrm{Black~curve}), \frac{1}{6}~(\mathrm{Blue~curve}), \frac{2}{3}~(\mathrm{Magenta~curve}), 1~(\mathrm{Green~curve}), 1.56~(\mathrm{Red~curve})$ respectively. Moreover the yellow curve in the right plot is the BBN temperature $\sim 10^{-2}\mathrm{GeV}$.}
\label{plot-R1}
\end{center}
\end{figure}

\begin{itemize}
 \item \underline{\boldmath{Set-1: $\lambda = -0.005$ $\eta = 1$ and $V_0 = 10^{-12}$}}: For this set, we show the required plots in Fig.[\ref{plot-R1}], see $N_\mathrm{re}$ vs. $N_\mathrm{f}$ in the left plot and $T_\mathrm{re}$ vs. $N_\mathrm{f}$ in the right plot. Fig.[\ref{plot-R1}] clearly demonstrates that $w_\mathrm{eff} > \frac{1}{3}$ allows $N_\mathrm{f} \gtrsim 54.30$, otherwise the reheating e-fold number becomes negative and thus is unphysical. On other hand, for $w_\mathrm{eff} < \frac{1}{3}$, the inflationary e-fold number should be $N_\mathrm{f} \lesssim 54.30$ to make the the $N_\mathrm{re}$ positive. Taking these constraints into account, we plot $T_\mathrm{re}$ vs. $N_\mathrm{f}$ for the aforementioned values of $w_\mathrm{eff}$, which depicts that the $T_\mathrm{re}$ remains less than the BBN temperature ($T_\mathrm{BBN} \sim 10^{-2}\mathrm{GeV}$) for such wide ranges of $w_\mathrm{eff}$. Thus combining the inflation and reheating constraints, what we find is following:
 \begin{eqnarray}
  w_\mathrm{eff} < \frac{1}{3}~~~~\mathrm{or}~~~m > \frac{1}{2}~~~~~~~~\Longrightarrow~~~~N_\mathrm{f} = [50,54.30]~~;\label{et-12a}
  \end{eqnarray}
  and
  \begin{eqnarray}
  \frac{1}{3} < w_\mathrm{eff} < 1.56~~~~\mathrm{or}~~~0.26 < m < \frac{1}{2}~~~~~~~~\Longrightarrow~~~~N_\mathrm{f} = [54.30,60]~~,
  \label{et-12b}
 \end{eqnarray}
 respectively, where we may recall that $w_\mathrm{eff} = -1 + \frac{2}{3m}$, see Eq.(\ref{reh-5}).

\end{itemize}

 \begin{figure}[!h]
\begin{center}
\centering
\includegraphics[width=3.0in,height=2.0in]{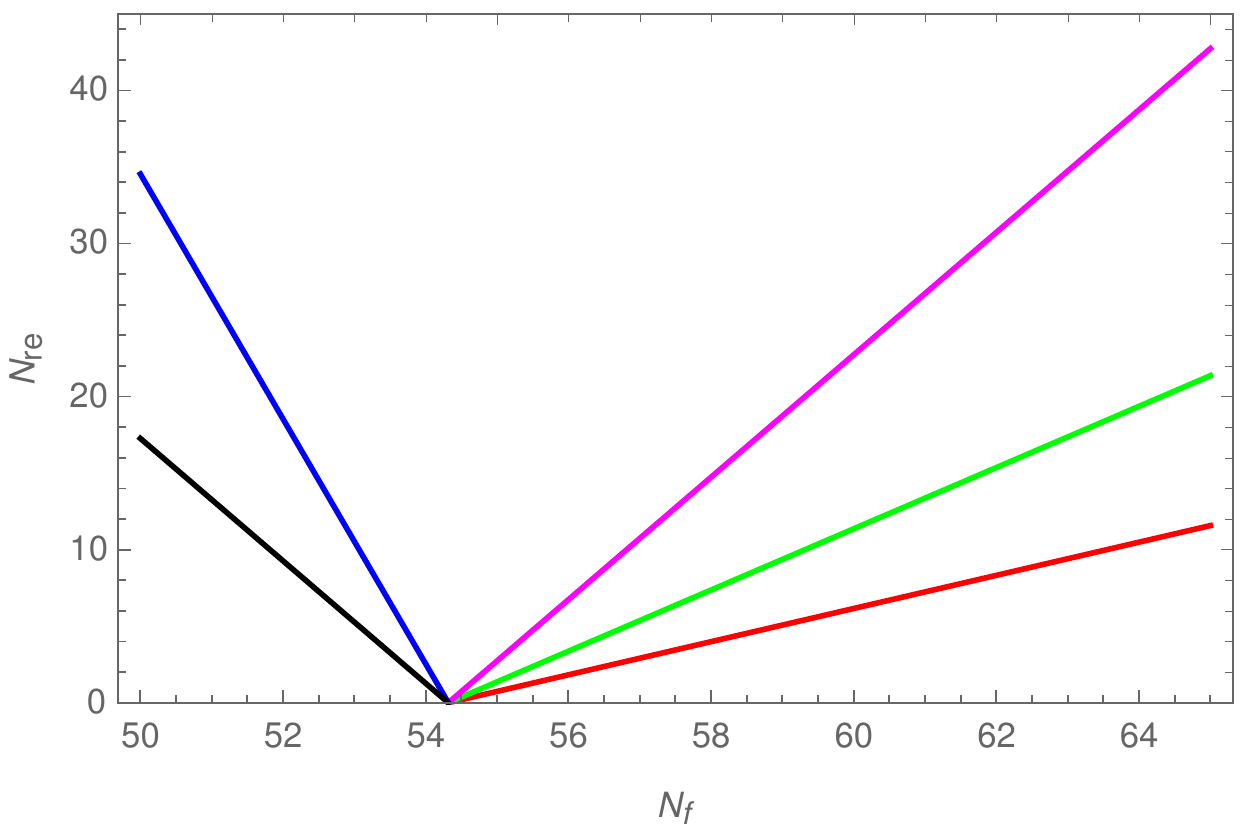}
\includegraphics[width=3.0in,height=2.0in]{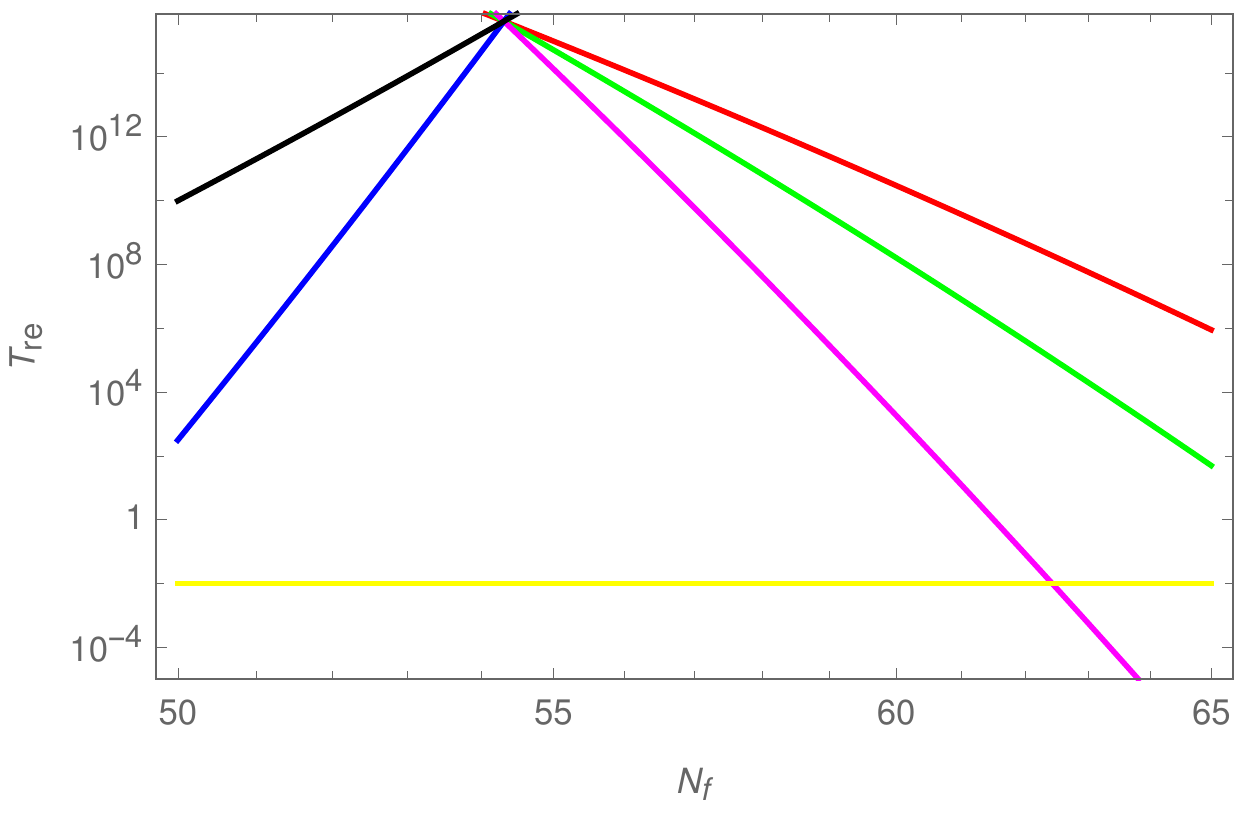}
\caption{{\color{blue}Left Plot}: $N_\mathrm{re}$ vs. $N_\mathrm{f}$; {\color{blue}Right Plot}: $T_\mathrm{re}$ (in GeV unit) vs. $N_\mathrm{f}$ for various values of $w_\mathrm{eff}$. In both the plots, we consider $\lambda = -0.001$ $\eta = 1$ and $V_0 = 10^{-12}$ (in Planck units). The reheating EoS parameter is taken as $w_\mathrm{eff} = 0 ~(\mathrm{Black~curve}), \frac{1}{6}~(\mathrm{Blue~curve}), \frac{2}{3}~(\mathrm{Magenta~curve}), 1~(\mathrm{Green~curve}), 1.56~(\mathrm{Red~curve})$ respectively. Moreover the yellow curve in the right plot is the BBN temperature $\sim 10^{-2}\mathrm{GeV}$.}
\label{plot-R2}
\end{center}
\end{figure}

\begin{itemize}
 \item \underline{\boldmath{Set-2: $\lambda = -0.001$ $\eta = 1$ and $V_0 = 10^{-12}$}}: For this set, the required plots are showed in Fig.[\ref{plot-R2}] where the left and the right plots depicts $N_\mathrm{re}$ vs. $N_\mathrm{f}$ and $T_\mathrm{re}$ vs. $N_\mathrm{f}$ respectively. Again in this case, in order to ensure a positive $N_\mathrm{re}$; the inflationary e-fold number should lie within $N_\mathrm{f} = [54.32,65]$ for $w_\mathrm{eff} > \frac{1}{3}$, while $w_\mathrm{eff} < \frac{1}{3}$ demands $N_\mathrm{f} = [50,54.32]$. However the reheating temperature, for this set, may go below the BBN temperature for some of $w_\mathrm{eff} > \frac{1}{3}$ values. In particular, for $w_\mathrm{eff} = \frac{2}{3}$, the reheating temperature gets below of BBN temperature when the inflationary e-fold number is given by $N_\mathrm{f} > 62$ -- this clearly reveals that the allowed range of $N_\mathrm{f}$ in case of $w_\mathrm{eff} = \frac{2}{3}$ should be $N_\mathrm{f} = [54.32,62]$. It turns out that $w_\mathrm{eff} > 0.83$ allows $N_\mathrm{f} = [54.3,65]$, while for $\frac{1}{3} < w_\mathrm{eff} < 0.83$, the $N_\mathrm{f}$ gets further constrained which may be determined from Eq.(\ref{et-11}). Thus as a whole,
 \begin{eqnarray}
  w_\mathrm{eff} < \frac{1}{3}~~~~\mathrm{or}~~~m > \frac{1}{2}~~~~~~~~\Longrightarrow~~~~N_\mathrm{f} = [50,54.32]~~;\label{et-13a}
  \end{eqnarray}
  and
  \begin{eqnarray}
  0.83< w_\mathrm{eff} < 1.56~~~~~~~\mathrm{or}~~~0.26 < m < 0.365~~~~~~~~\Longrightarrow~~~~N_\mathrm{f} = [54.32,65]~~.\label{et-13}
 \end{eqnarray}
As we have just mentioned that the allowed range of $N_\mathrm{f}$ for $\frac{1}{3} < w_\mathrm{eff} < 0.83$ can be determined with specific value of $w_\mathrm{eff}$ by using Eq.(\ref{et-11}).

\end{itemize}

As a whole, we put the constraints on $N_\mathrm{f}$ (coming from both the inflation and reheating phenomenology) in the following Table[\ref{Table-0}].

\begin{table}[h]
  \centering
 {%
  \begin{tabular}{|c|c|c|}
   \hline
    Model parameters & Reheating EoS parameter & Constraint on $N_\mathrm{f}$\\

   \hline
  \hline
   (1) $\lambda = -0.005$, $\eta = 1$, $V_0 = 10^{-12}$ (in Planck units) & (a) $w_\mathrm{eff} < \frac{1}{3}$ & $N_\mathrm{f} = [50,54.30]$\\
   \hline
     & (b) $\frac{1}{3} < w_\mathrm{eff} < 1.56$ & $N_\mathrm{f} = [54.30,60]$\\
     \hline
    (2) $\lambda = -0.001$, $\eta = 1$, $V_0 = 10^{-12}$ (in Planck units) & (a) $w_\mathrm{eff} < \frac{1}{3}$ & $N_\mathrm{f} = [50,54.32]$\\
   \hline
     & (b) $0.83 < w_\mathrm{eff} < 1.56$ & $N_\mathrm{f} = [54.32,65]$\\
   \hline
   \hline
  \end{tabular}%
 }
  \caption{Constraints on $N_\mathrm{f}$ (coming from both the inflation and reheating phenomenology) for two different sets of model parameters.}
  \label{Table-0}
 \end{table}

Thus we may notice that the inflationary e-fold number in the present context gets further constrained by the input of the reheating stage.

\subsection*{Possibility of instantaneous reheating}
In the case of instantaneous reheating, the scalar field instantaneously decays to radiation $immediately~after~the~end~of~inflation$. Therefore the reheating e-fold number is zero, i.e $N_\mathrm{re} = 0$, and the reheating temperature from Eq.(\ref{et-4}) comes as,
\begin{eqnarray}
 T_\mathrm{re} = H_\mathrm{i}\left(\frac{43}{11g_\mathrm{re}}\right)^{1/3}\left(\frac{T_0}{k/a_0}\right)e^{-N_\mathrm{f}}~~.
 \label{pos-1}
\end{eqnarray}
Moreover, due to $N_\mathrm{re} = 0$, Eq.(\ref{et-10}) immediately leads to,
\begin{eqnarray}
 N_\mathrm{f} = 61.6 - \frac{1}{4}\ln{\left(\frac{9e^{\lambda\left(2\phi_\mathrm{i} - \phi_\mathrm{f}\right)}}{2V_0}\right)}~~.
 \label{pos-2}
\end{eqnarray}
Therefore the instantaneous reheating is allowed if the inflationary e-fold number satisfies the above equation. Thus in order to examine the possibility of instantaneous reheating, we need to solve $N_\mathrm{f}$ from Eq.(\ref{pos-2}), and then check whether the solution of $N_\mathrm{f}$ obtained from Eq.(\ref{pos-2}) leads to viable $n_s$ and $r$ with respect to the Planck data. For this purpose, we use the aforementioned sets of model parameters and determine $N_\mathrm{f}$, $(n_s, r)$ and $T_\mathrm{re}$ -- this is depicted in the following Table [\ref{Table-1}].

\begin{table}[h]
  \centering
 {%
  \begin{tabular}{|c|c|c|c|}
   \hline
    Model parameters & $N_\mathrm{f}$ from Eq.(\ref{pos-2}) & ($n_s$, $r$) from Eq.(\ref{observable-3}) & $T_\mathrm{re}$ (GeV) from Eq.(\ref{pos-1})\\

   \hline
  \hline
   $(1) \lambda = -0.005$, $\eta = 1$, $V_0 = 10^{-12}$ (in Planck units) & $54.30$ & (0.9685, 0.003) & $4.10\times10^{15}$\\
   \hline
   $(2) \lambda = -0.001$, $\eta = 1$, $V_0 = 10^{-12}$ (in Planck units) & $54.32$ & (0.9648, 0.002) & $4.17\times10^{15}$\\
   \hline
   \hline
  \end{tabular}%
 }
  \caption{$N_\mathrm{f}$, $(n_s, r)$ and $T_\mathrm{re}$ for two different sets of model parameters in the case of instantaneous reheating.}
  \label{Table-1}
 \end{table}

 Table[\ref{Table-1}] clearly demonstrates that th instantaneous reheating in the present context of scalar-Einstein-GB gravity is allowed as it leads to viable $(n_s, r)$ and also the reheating temperature is safe from the BBN temperature.

 \section{Full form of scalar potential and GB coupling function: Numerical solution}\label{sec-complete}

 In this section, we establish the complete forms of scalar potential ($V(\phi)$) and GB coupling function ($\xi(\phi)$) during the entire epoch, in particular, such complete forms of $V(\phi)$ and $\xi(\phi)$ smoothly transits from the inflation to the reheating stage. Recall from Eq.(\ref{reh-pot}) that the scalar potential during inflation and during reheating are given by,
 \begin{eqnarray}
  V_\mathrm{I}(\phi)&=&V_0~e^{-\lambda\phi}~~,\nonumber\\
  V_\mathrm{R}(\phi)&=&V_\mathrm{1}~e^{2\left(\phi - \phi_\mathrm{s}\right)/\phi_\mathrm{0}}~~,
  \label{comp-1}
 \end{eqnarray}
respectively, where $V_\mathrm{1} = 5c_1m^2 - c_1m + 3m^2 - m$ (see Eq.(\ref{reh-7})). Based on the above forms of $V_\mathrm{I}(\phi)$ and $V_\mathrm{R}(\phi)$, we may write the full forms of scalar field potential as,
 \begin{eqnarray}
  V(\phi)&=&\frac{1}{2}\left[V_\mathrm{I}(\phi)\left\{1 + \mathrm{tanh}\left[p\left(\phi - \phi_\mathrm{f}\right)\right]\right\} + V_\mathrm{R}(\phi)\left\{1 - \mathrm{tanh}\left[p\left(\phi - \phi_\mathrm{f}\right)\right]\right\}\right]\nonumber\\
  &=&\frac{1}{2}\left[V_0~e^{-\lambda\phi}\left\{1 + \mathrm{tanh}\left[p\left(\phi - \phi_\mathrm{f}\right)\right]\right\} + V_\mathrm{1}~e^{2\left(\phi - \phi_\mathrm{s}\right)/\phi_\mathrm{0}}\left\{1 - \mathrm{tanh}\left[p\left(\phi - \phi_\mathrm{f}\right)\right]\right\}\right]~~,
  \label{comp-2}
 \end{eqnarray}
with $p > 1$ be a controlling parameter that controls the smooth transition of the scalar potential from inflation to reheating, and $\phi_\mathrm{f}$ denotes the scalar field at the end of inflation. Eq.(\ref{comp-2}) depicts that for $\phi > \phi_\mathrm{f}$, i.e during the inflation (recall $\phi_\mathrm{i} > \phi_\mathrm{f}$), the $V(\phi)$ behaves as $V(\phi) \approx V_\mathrm{I}(\phi)$; and for $\phi < \phi_\mathrm{f}$, i.e during the reheating, the $V(\phi) \approx V_\mathrm{R}(\phi)$. Moreover, for a smooth transition of the scalar potential, we take $V_\mathrm{I}(\phi_\mathrm{f}) = V_\mathrm{R}(\phi_\mathrm{f})$ (i.e the inflationary scalar potential and the reheating scalar potential are equal at the junction of inflation-to-reheating),
 \begin{eqnarray}
  V_\mathrm{I}(\phi_\mathrm{f})&=&V_\mathrm{R}(\phi_\mathrm{f})\nonumber\\
  \Longrightarrow V_\mathrm{0}~e^{-\lambda\phi_\mathrm{f}}&=&V_\mathrm{1}~e^{2\left(\phi_\mathrm{f} - \phi_\mathrm{s}\right)/\phi_\mathrm{0}}~~,
  \label{comp-3}
 \end{eqnarray}
which immediately leads to,
 \begin{eqnarray}
  \phi_\mathrm{s} = \left(\frac{\phi_\mathrm{0}}{2}\right)\ln{\left[\left(\frac{V_\mathrm{1}}{V_\mathrm{0}}\right)\mathrm{exp}\left\{\left(\frac{2}{\phi_\mathrm{0}} + \lambda\right)\phi_\mathrm{f}\right\}\right]}~~.
  \label{comp-4}
 \end{eqnarray}
Here $\phi_\mathrm{0} = \sqrt{2m\left(c_1m + c_1 + 1\right)}$, see Eq.(\ref{reh-7}). The above equation justifies the presence of $\phi_\mathrm{s}$ in the form of $V_\mathrm{R}(\phi)$. By getting an expression like Eq.(\ref{comp-4}), the parameter $\phi_\mathrm{s}$ actually helps to smoothly connect the $V_\mathrm{I}(\phi)$ to $V_\mathrm{R}(\phi)$ at the junction of inflation-to-reheating at $\phi = \phi_\mathrm{f}$. Following the same analysis, we write the complete form of the GB coupling function as,
 \begin{eqnarray}
  \xi(\phi)&=&\frac{1}{2}\left[\xi_\mathrm{I}(\phi)\left\{1 + \mathrm{tanh}\left[p\left(\phi - \phi_\mathrm{f}\right)\right]\right\} + \xi_\mathrm{R}(\phi)\left\{1 - \mathrm{tanh}\left[p\left(\phi - \phi_\mathrm{f}\right)\right]\right\}\right]\nonumber\\
  &=&\frac{1}{2}\left[\xi_\mathrm{0}~e^{-\eta\phi}\left\{1 + \mathrm{tanh}\left[p\left(\phi - \phi_\mathrm{f}\right)\right]\right\} + \xi_\mathrm{1}~e^{-2\left(\phi - \phi_\mathrm{s}\right)/\phi_\mathrm{0}}\left\{1 - \mathrm{tanh}\left[p\left(\phi - \phi_\mathrm{f}\right)\right]\right\}\right]~~,
  \label{comp-5}
 \end{eqnarray}
where we use the expressions of $\xi_\mathrm{I}(\phi)$ and $\xi_\mathrm{R}(\phi)$ from Eq.(\ref{reh-7}).

\begin{figure}[!h]
\begin{center}
\centering
\includegraphics[width=3.0in,height=2.0in]{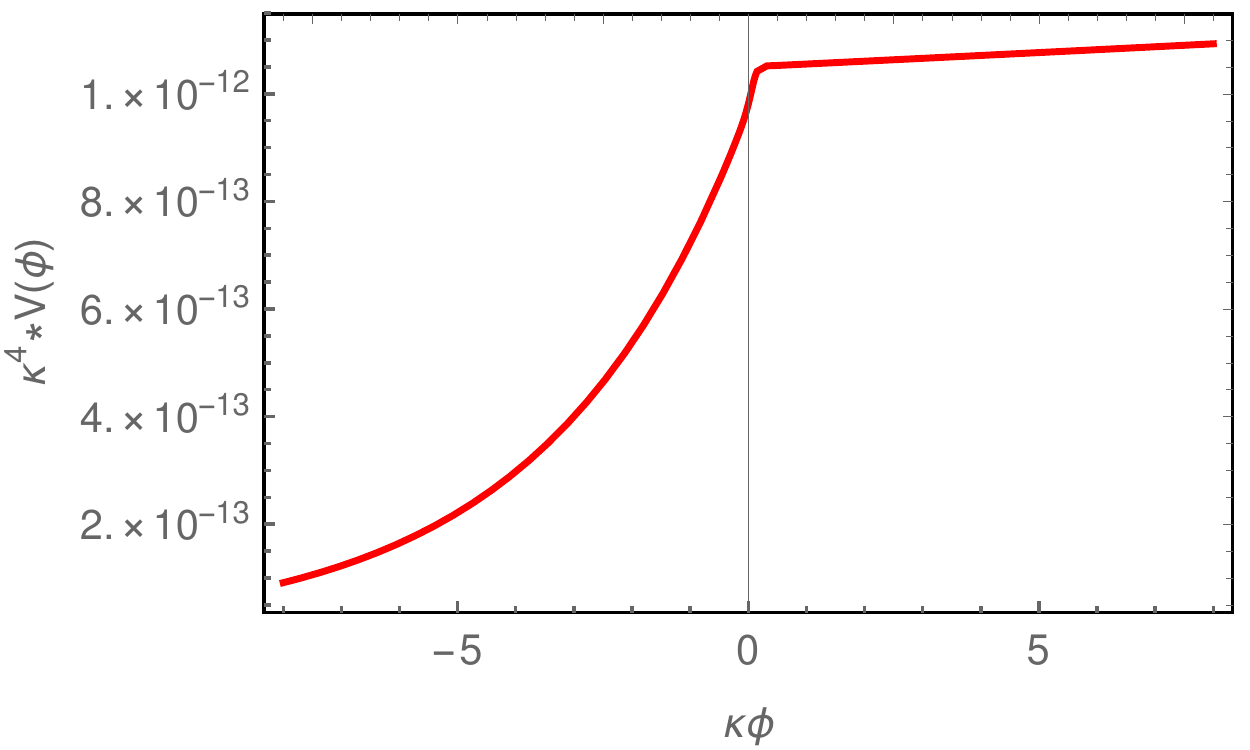}
\includegraphics[width=3.0in,height=2.0in]{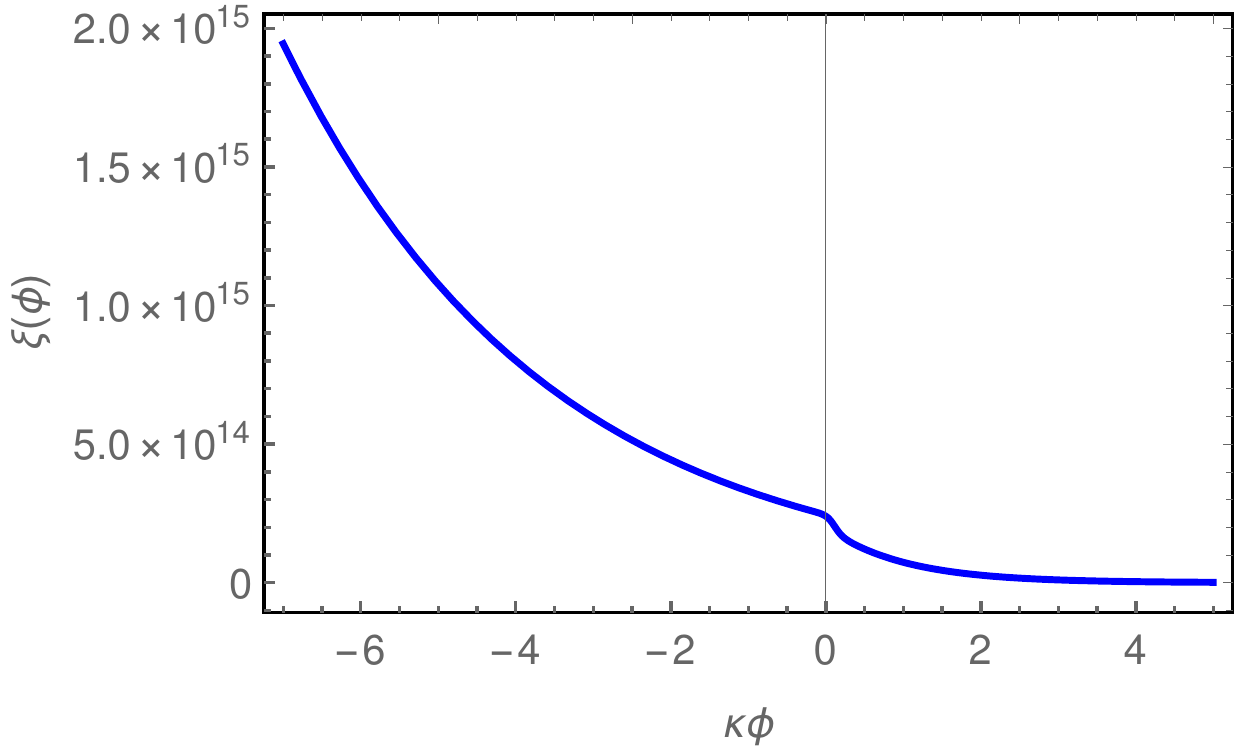}
\caption{{\color{blue}Left Plot}: $\kappa^4V(\phi)$ vs. $\kappa\phi$ from Eq.(\ref{comp-3}); {\color{blue}Right Plot}: $\xi(\phi)$ vs. $\kappa\phi$ from Eq.(\ref{comp-5}). Recall that $\kappa^{-1} = \frac{1}{\sqrt{8\pi G}} \approx 10^{19}\mathrm{GeV}$ (with $G$ being the Newton's gravitational constant). In both the plots, we use the following set of parameters: $\lambda = -0.005$, $\eta = 1$, $\kappa^4V_\mathrm{0} = 10^{-12}$, $\kappa^4V_\mathrm{0}\xi_\mathrm{0} = 1$ and $m = \frac{2}{3}$ (i.e for $w_\mathrm{eff} = -1 + \frac{2}{3m} = 0$).}
\label{plot-comp1}
\end{center}
\end{figure}

\begin{figure}[!h]
\begin{center}
\centering
\includegraphics[width=3.0in,height=2.0in]{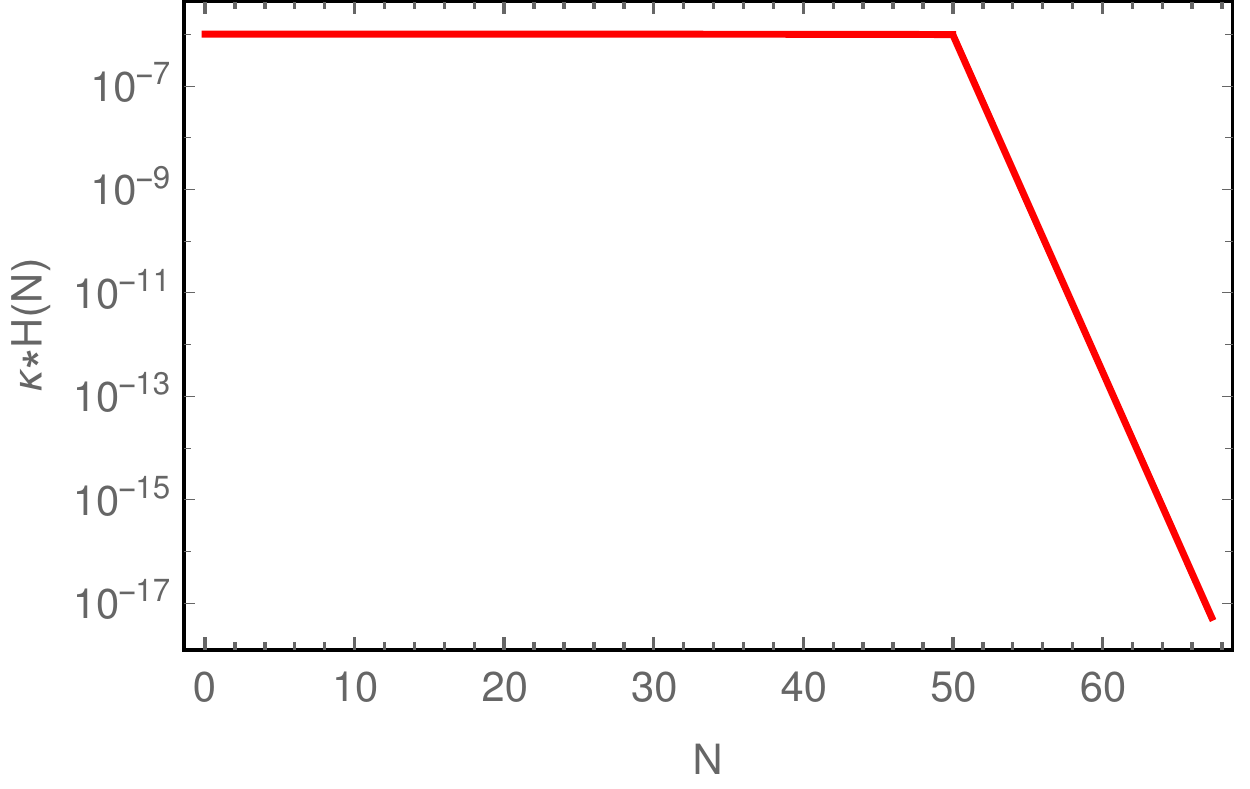}
\includegraphics[width=3.0in,height=2.0in]{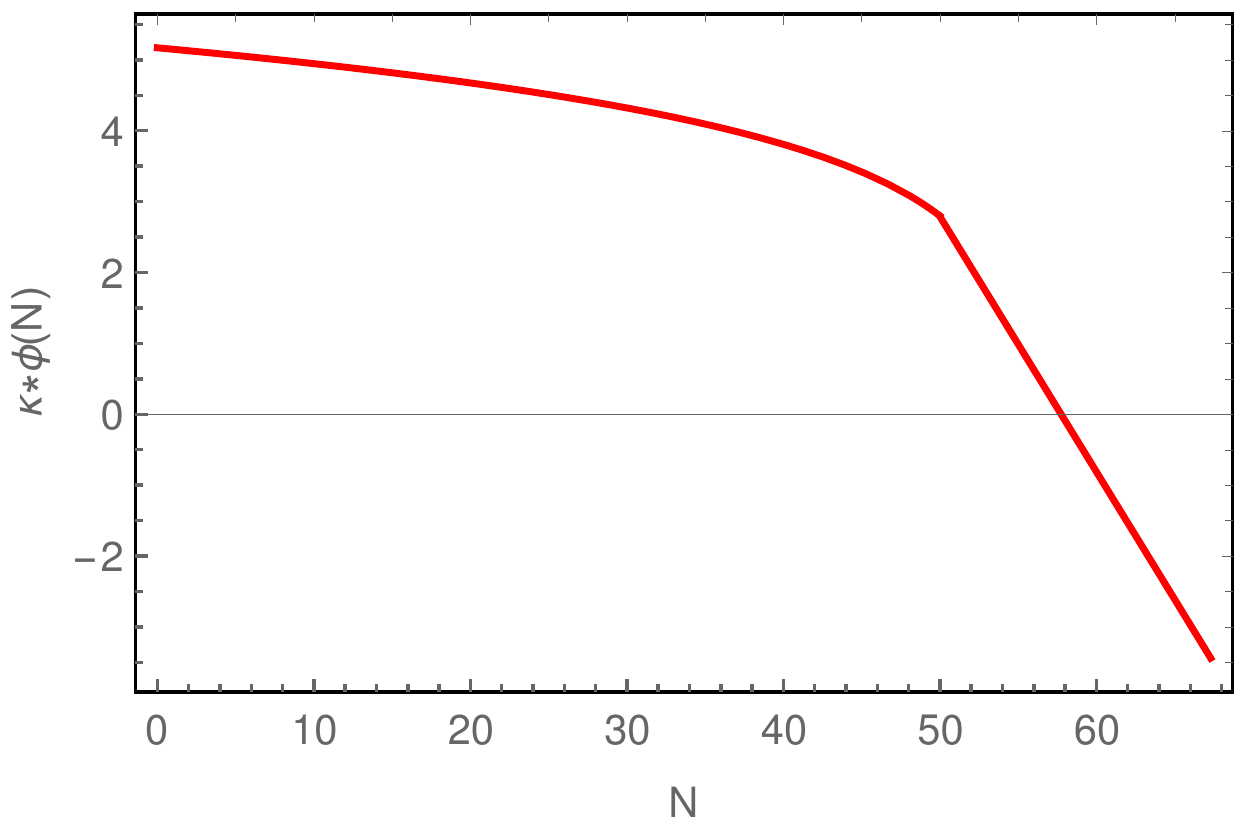}
\caption{{\color{blue}Left Plot}: $\kappa H(N)$ (in logarithmic scale) vs. $N$; {\color{blue}Right Plot}: $\kappa\phi(N)$ vs. $N$. Similar to the Fig.[\ref{plot-comp1}], $\kappa^{-1} = \frac{1}{\sqrt{8\pi G}} \approx 10^{19}\mathrm{GeV}$ (with $G$ being the Newton's gravitational constant). In both the plots, we use the following set of parameters: $\lambda = -0.005$, $\eta = 1$, $\kappa^4V_\mathrm{0} = 10^{-12}$, $\kappa^4V_\mathrm{0}\xi_\mathrm{0} = 1$ and $m = \frac{2}{3}$. For such set of parameters, the inflationary e-fold number gets constrained by $N_\mathrm{f} = [50,54.30]$; thus we safely take $N_\mathrm{f} = 50$ in the plots. Moreover the reheating e-fold number is $N_\mathrm{re} = 17.3$. Thus the inflation lasts for $0 < N \leq N_\mathrm{f} = 50$ and the reheating duration is given by $N_\mathrm{f} < N \leq N_\mathrm{f} + N_\mathrm{re} = 67.3$.}
\label{plot-comp2}
\end{center}
\end{figure}

By using Eq.(\ref{comp-3}) and Eq.(\ref{comp-5}), we plot $V(\phi)$ and $\xi(\phi)$ respectively in Fig.[\ref{plot-comp1}] for the following set of parameters which we have used earlier: $\lambda = -0.005$, $\eta = 1$, $\kappa^4V_\mathrm{0} = 10^{-12}$, $\kappa^4V_\mathrm{0}\xi_\mathrm{0} = 1$ and $m = \frac{2}{3}$ (i.e for $w_\mathrm{eff} = -1 + \frac{2}{3m} = 0$). This set of parameters along with Eq.(\ref{end phi}) lead to $\kappa\phi_\mathrm{f} = 0.98$ i.e. when the inflation ends. Consequently, by using the full forms of $V(\phi)$ and $\xi(\phi)$, we numerically solve the Hubble parameter and the scalar field (with respect to e-fold number) from the inflation to the reheating stage, which are depicted in Fig.[\ref{plot-comp2}]. In particular, we give the plots for $\kappa H(N)$ and $\kappa\phi(N)$, with $\kappa^{-1} = \frac{1}{\sqrt{8\pi G}} \approx 10^{19}\mathrm{GeV}$ (with $G$ being the Newton's gravitational constant). The initial conditions for such numerical calculations are taken from the inflationary analytic solutions of $\phi(N)$ and $H(N)$ from Eq.(\ref{e-fold-4}) and Eq.(\ref{Hubble-evolution}) respectively, i.e.,
\begin{eqnarray}
  \phi(N=0) = \frac{1}{\left(\lambda + \eta\right)}
  \ln{\left[\frac{V_0\xi_0\eta}{3\lambda}\left\{2\mathrm{e}^{-\lambda\left(\lambda + \eta\right)N_\mathrm{f}}
  \left(\lambda\sqrt{\left(\lambda^2 + 4\right)} - \lambda^2 - 2\right) + 4\right\}\right]}~~,
  \label{NN-1}
 \end{eqnarray}
 and
 \begin{eqnarray}
  H(N=0) = \left(\frac{V_0}{3}\right)^{\frac{1}{2}}
  \left[\frac{V_0\xi_0\eta}{3\lambda}\left\{2\mathrm{e}^{-\lambda\left(\lambda + \eta\right)N_\mathrm{f}}
  \left(\lambda\sqrt{\left(\lambda^2 + 4\right)} - \lambda^2 - 2\right) + 4\right\}\right]^{\frac{-\lambda}{2\left(\lambda + \eta\right)}}~~.
  \label{NN-2}
 \end{eqnarray}
Recall from Table[\ref{Table-0}] that for aforementioned parameter values, the inflationary e-fold number gets constrained by $N_\mathrm{f} = [50,54.30]$; thus we safely take $N_\mathrm{f} = 50$ in the plots. Moreover Fig.[\ref{plot-R1}] indicates that $N_\mathrm{re} = 17.3$ for $m = \frac{2}{3}$. Thus in the Fig.[\ref{plot-comp2}], the inflation lasts for $0 < N \leq N_\mathrm{f} = 50$ and the reheating duration is given by $N_\mathrm{f} < N \leq N_\mathrm{f} + N_\mathrm{re} = 67.3$.

Fig.[\ref{plot-comp2}] clearly demonstrates that the Hubble parameter remains almost constant during the inflation (i.e for $0 < N \leq N_\mathrm{f} = 50$) and evolves through a power law form during the reheating stage (i.e for $N_\mathrm{f} < N \leq N_\mathrm{f} + N_\mathrm{re} = 67.3$). Thus the numerical solutions match with the analytic ones in the respective stages. This in turn ensures the validity of the numerical solutions.

\section{Conclusion}

The present work devotes to investigate the inflation and reheating phenomenology in the context of scalar-Einstein-GB theory where the scalar field under consideration non-minimally coupes with the GB curvature term. The GB gravity theory is well motivated due to the fact that the gravitational field equations are of second order (with respect to time) and thus the theory is free from any Ostragradsky instability. The scalar potential and the GB coupling function are considered to be of exponential types, which allows analytic solutions of the field variables (i.e of the Hubble parameter and the scalar field) during inflation as well as during reheating era. Regarding the inflation, the findings can be summarized as follows: (1) the inflation begins with a quasi de-Sitter phase and has an exit at finite e-fold number ($N_\mathrm{f}$) that lies within $N_\mathrm{f} = [50, 60]$ or $N_\mathrm{f} = [50, 65]$ suitable for resolving the horizon and flatness problems; (2) the primordial scalar and tensor perturbations prove to be stable and are free from gradient instability; (3) the inflationary parameters, in particular the amplitude of the curvature perturbation as well as its tilt and the tensor to scalar ratio turn out to be simultaneously compatible for suitable values of the model parameters which also leads to the inflationary energy scale $\sim 10^{13}\mathrm{GeV}$. After the inflation ends, the scalar field decays to radiation energy density with a constant decay width (symbolized by $\Gamma$). Actually after the end of inflation, the Hubble parameter is much larger than the decay width and hence the decay of the scalar field is negligible with respect to the Hubble expansion at the beginning of the reheating era. However the Hubble parameter continually decreases with the universe's expansion and eventually it becomes comparable with the decay width, i.e $H \sim \Gamma$, then the scalar field decays to radiation and $H = \Gamma$ represents the end of inflation. For our considered scalar potential and the GB coupling function, the present model leads to a power law solution of the Hubble parameter and a logarithmic solution of the scalar field during the reheating era, where the exponent of the Hubble parameter determines the effective EoS parameter ($w_\mathrm{eff}$) during the same. Here it deserves mentioning that the $w_\mathrm{eff}$ in the present context is contributed from the canonical part of the scalar field as well as from the coupling between the GB term and the scalar field, owing to which, the GB coupling has significant effects on $w_\mathrm{eff}$. The stability of the reheating dynamics is examined by dynamical analysis which ensures that the reheating EoS parameter can go beyond unity and reaches up-to the value of $\mathrm{max}(w_\mathrm{eff}) = 1.56$. This is unlike to canonical scalar-tensor theory (where the GB coupling is absent) where the stability of reheating dynamics demands that the reheating EoS parameter must be less than unity. On contrary, in the context of scalar-Einsten-GB gravity, the presence of GB coupling function ensures the stability of the background reheating dynamics even for $w_\mathrm{eff} > 1$, in particular for $w_\mathrm{eff} < 1.56$. This may have important consequences on primordial gravitational waves' (PGWs') spectrum (observed today) as the reheating EoS parameter is expected to play an important role on enhancing the PGWs' amplitude \cite{Haque:2021dha}. The e-fold number of the reheating era ($N_\mathrm{re}$) and the reheating temperature ($T_\mathrm{re}$) are determined in terms of inflationary quantities and $w_\mathrm{eff}$. By the inputs of the reheating stage, in particular by $N_\mathrm{re} > 0$ and $T_\mathrm{re} < T_\mathrm{BBN} \sim 10^{-2}\mathrm{GeV}$, the inflationary e-fold number (i.e $N_\mathrm{f}$) gets further constrained, and such constraints coming from both the inflation and reheating phenomenology are critically examined (see Fig.[\ref{plot-R1}], Fig.[\ref{plot-R2}] and Table[\ref{Table-0}]). Finally we establish complete forms of scalar potential ($V(\phi)$) and GB coupling function ($\xi(\phi)$) that smoothly transits from the inflation to the reheating stage, and consequently, numerically solve the Hubble parameter and the scalar field during the entire epoch. The numerical solution depicts that the Hubble parameter follows a quasi de-Sitter evolution and a power law type solution during the inflation and the reheating era respectively. This indeed confirms that the numerical solutions match with the analytic ones during the respective eras. After the end of reheating, the GB term becomes a surface term in the four dimensional gravitational action due to the fact that the whole scalar field energy density decays to radiation, and thus the gravitational action contains the usual Einstein-Hilbert term with radiation as matter component. This results to the standard radiation era of the universe.

\end{document}